\documentclass[article]{JHEP3}

\usepackage{amsmath,epsfig}

\usepackage{amssymb,amsfonts,cite}

\title{Closed Strings in Misner Space:\\Cosmological Production of
  Winding Strings }
\preprint{\hepth{0405126}\\LPTHE-04-04\\WIS/20/04-JUL-DPP}

\author{M. Berkooz\footnote{Incumbent of the Recanati career
development chair for energy research}\\
Weizmann Institute of Science,\\ Rehovot 76100,
Israel\\{\tt E-mail:
berkooz@wisemail.weizmann.ac.il}}

\author{B. Pioline\\
LPTHE, Universit\'es Paris 6 et 7, 4 place Jussieu, \\
75252 Paris cedex 05, France\\{\tt E-mail:
pioline@lpthe.jussieu.fr}}

\author{M. Rozali\\
Department of Physics and Astronomy, 6224 Agricultural Rd.\\
University of British Columbia, Vancouver, BC V6T 1Z1, Canada\\
{\tt E-mail: rozali@physics.ubc.ca}}

\abstract{Misner space, also known as the
Lorentzian orbifold $\Real^{1,1}/boost$,
is one of the simplest examples of a cosmological
singularity in string theory. In this work, the
study of weakly coupled closed strings on this space is pursued
in several directions:
(i) physical states in the twisted sectors
are found to come in two kinds:
{\it short strings}, which wind along the compact
space-like direction
in the cosmological (Milne) regions,
 and {\it long strings},
which wind along the compact time-like direction in the
(Rindler) whiskers. The latter can be viewed as
infinitely long static open strings, stretching
from Rindler infinity to a finite radius and
folding back onto themselves. (ii) As in the
Schwinger effect, tunneling between these states
corresponds to local pair production of winding
strings.  The tunneling rate approaches unity as
the winding number $w$ gets large, as a consequence
of the singular geometry. (iii) The one-loop string amplitude
has singularities on the moduli space,  associated
to periodic closed string trajectories in Euclidean
time. In the untwisted sector, they can be traced
to the combined existence of CTCs and Regge trajectories 
in the spectrum. In the twisted sectors, they
indicate pair production of winding strings.
(iv) At a classical level and in sufficiently low
dimension, the condensation of winding strings can
indeed lead to a bounce, although the required
initial conditions are not compatible with Misner
geometry at early times. (v) The semi-classical
analysis of winding string pair creation can be generalized to
more general (off-shell) geometries. We show that a
regular geometry regularizes the divergence at
large winding number. }

\makeatletter
\renewcommand{\subsubsection}{\@startsection{subsubsection}{3}{0mm}{-\baselineskip}{0.5\baselineskip}{\normalfont\normalsize\it}}
\makeatother

\newcommand{\pa}{\partial}

\newcommand{\p}{\partial}

\newcommand{\nn}{\nonumber}
\newcommand{\eps}{\epsilon}
\newcommand{\Real}{\mathbb{R}}
\newcommand{\Zint}{\mathbb{Z}}
\newcommand{\Nint}{\mathbb{N}}

\newcommand{\sgn}{\mbox{sgn}}

\def\bea{\begin{eqnarray}}
\def\eea{\end{eqnarray}}
\def\be{\begin{equation}}
\def\ee{\end{equation}}
\def\ba{\begin{align}}
\def\ea{\end{align}}
\def\bse{\begin{subequations}}
\def\ese{\end{subequations}}
\def\bi{\begin{itemize}}
\def\ei{\end{itemize}}

\def\a{\alpha}
\def\talpha{\tilde\alpha}
\def\teps{\tilde\epsilon}

\def\tk{\tilde k}
\def\tR{\tilde R}
\def\ta{\tilde \alpha}
\def\tM{\tilde M}
\def\1F1{{}_1\!F_1}
\def\2F0{{}_2\!F_0}

\begin{document}
\maketitle \setcounter{tocdepth}{2}
\tableofcontents

\section{Introduction}

In spite of its great power in inspiring new
cosmological scenarios, string theory in
cosmological backgrounds remains a mostly uncharted
territory. One may hope that inherently stringy
phenomena play a r\^ole in early universe cosmology,
possibly alleviating some of the problems facing an
effective field theory approach to that subject.
 For example, in perturbative string theory,
 the exponentially growing density of states and the
existence of topologically stable excitations are
expected to lead to strong departures from ordinary
FRW cosmology near the string scale, and possibly
to observable stringy signatures
\cite{Brandenberger:1988aj}. This context
requires the development of new tools for
investigating string theory in a dynamical setting.

 Cosmological
production of winding and  excited strings, as well
as other solitons, would perhaps be best described
in the framework of a field theory of closed
strings, a sorely missing item from the string
theorist's toolbox. It is thus important to see how
far the ordinary, first quantized world-sheet
description of strings can be pushed to address
these questions (see \cite{Lawrence,Gubser,Kabat} for
recent progress).
We will see that application of
this formalism, even in the simple context studied
here, is subtle and requires extending and
generalizing the usual rules of string perturbation
theory.

Following  earlier work \cite{Nappi:1992kv,Khoury:2001bz,Nekrasov:2002kf,
Elitzur:2002rt,lms,Cornalba:2002fi,Craps:2002ii,Fabinger:2002kr,Berkooz:2002je,
Pioline:2003bs,Craps:2003ai} on related models, we
study here a simple model of a cosmological
singularity, known as the Misner (or Milne)
Universe \cite{Misner}. Being the orbifold of flat
Minkowski space $\Real^{1,1}$ by a discrete boost,
the world-sheet theory is in principle solvable by
standard conformal field theory techniques. The
adjustments entailed by the Lorentzian signature of the background
may nevertheless be non-trivial, as indicated by the divergences
found in the tree-level scattering amplitudes of untwisted
states \cite{Berkooz:2002je},  possibly hinting at a
breakdown of perturbation theory  \cite{lms,Horowitz:2002mw}.
Whether Misner space inevitably collapses gravitationally
or rebounces to a new expanding epoch is still
a matter for debate.

In an earlier work \cite{Pioline:2003bs}, two of the present authors
focused on the twisted sector of the orbifold,
which for Euclidean  orbifolds was the key to
resolving   the singularity. By noting the analogy
with   charged particles in an electric field, it
was shown that there exists a delta-function
normalizable continuum of physical twisted states,
which may in principle be pair-produced by the
Schwinger mechanism \cite{Schwinger:nm}. Since the
energy of the produced winding strings scales with
the radius of the Milne Universe, hence acts like
a positive cosmological constant, it
is conceivable that the resulting transient inflation be sufficient
to halt the cosmological collapse. If so,
the instability towards black hole creation raised in
\cite{Horowitz:2002mw} could be evaded
and a smooth transition to an expanding region may take place,
although perturbation theory may not suffice to describe
this process consistently.

In this work, we continue our investigations of
closed strings in Misner space in several
directions. Some of the issues we raise and conclusions
we reach are expected to be generic in any dynamical background of
string theory.

We begin, in Section 2, by summarizing known
features about quantum fields in Misner space, and
its deformation known as Grant space
\cite{Grant:1992kj}, or the electric Melvin
universe \cite{Cornalba:2002fi}. We comment in particular on
subtleties arising when quantizing higher spin
fields in this background. We furthermore discuss the one-loop
energy-momentum tensor for free fields in such backgrounds,  
to be compared with the string theory answer in Section 4.

In Section 3 we discuss the quantization of twisted strings
in this background. First we clarify the
relation to the charged particle case, and give a
new classical interpretation of the winding states:
in addition to the strings winding around the Milne
direction, which will be dubbed ``short strings'',
there exist new classical configurations of ``long
strings'' living purely in the Rindler regions of
the Lorentzian orbifold; these objects are static
with respect to Rindler time, and correspond to
infinite strings extending from infinity to a
finite radius and back to Rindler infinity again.
In contrast, short strings extend from the Milne
into the Rindler regions, and have a finite length
proportional to their winding number. Quantum
mechanically, a long string coming from infinity
may tunnel near its turning point into a short
string, and escape to the Milne future region.
We expect that such ``long strings'' exist in
other examples locally equivalent to the Lorentzian
orbifold. We conclude Section 3 by discussing the second
quantization of the twisted sector modes. We
concentrate on the vacuum ambiguity in this sector,
and demonstrate that additional choices for twisted
sector vacua exist, which are "stringy" in the
sense that they treat the left and right movers
asymmetrically.

In Section 4 we investigate the target space interpretation
of the one-loop amplitude
for closed strings in Misner and Grant spaces. This
partition function was computed in
\cite{Nekrasov:2002kf,Cornalba:2002fi} and shown to
have various divergences in the bulk of the moduli
space.  We compare the zero-mode part of the
amplitude in the untwisted sector to the field theory answer, and
trace the divergences in the untwisted
sector to the existence of Regge trajectories with
arbitrarily high spins. Moreover, we revisit the divergences
in the twisted sectors, and relate them to periodic closed
string trajectories in imaginary proper time. While the one-loop
amplitude remains real, we find instabilities corresponding to
the spontaneous production of (short and long) winding strings.

Finally, in Section 5, we consider the effect of the production
of winding strings on the cosmological evolution,
at the classical level. We show that twisted mode
condensation may lead, at least with low
co-dimension, to  smooth bouncing geometries.
Self-consistently, in such putative smooth
geometries the infinite production rate of winding
strings gets regularized. The issue of the proper
implementation of  back-reaction in string
perturbation theory, possibly generalizing similar
attempts in the case of electric fields \cite{Cooper:1992hw},
 is very challenging, and merits
more work.

\section{Untwisted Fields in Misner Space}

In this Section we summarize some of the known
features about field theory in Misner space, as
well as their extension to string theory in the
untwisted sector. In particular we construct
wave functions for higher spin fields and note that
their  convergence requires regularization of the
orbifold action in a manner explained below. We then
briefly recall a deformation of Misner
space known as Grant space, or the electric Melvin
universe, and review the computation of the one-loop
energy-momentum tensor in such backgrounds, for
comparison with the string theory answer in Section 4.

\subsection{Classical Propagation in Misner Space}
Misner space was introduced long ago as a toy model
for the singularities of the Lorentzian
Taub-NUT space \cite{Misner}. It is defined as the
quotient of flat Minkowski space by a  boost,
$x^\pm \equiv x^\pm e^{\pm 2\pi \beta}$. Defining
coordinates
\bea x^\pm  &=& T e^{\pm \beta
\theta}/\sqrt{2}\ ,\quad x^+ x^->0
~\mbox{(Milne regions)} \label{xpm1} \\
x^\pm  &=& \pm r e^{\pm \beta \eta}/\sqrt{2}\
,\quad x^+ x^-<0 ~\mbox{(Rindler regions)} \label{xpm2}\eea
the
metric can be written as
\be ds^2 = -2dx^+ dx^- +
(dx^i)^2 = \left\{
\begin{matrix}
-dT^2 + \beta^2 T^2 d\theta^2\\
dr^2 - \beta^2 r^2 d\eta^2
\end{matrix}
\right\}  + (dx^i)^2
\ee
where the spatial direction $\theta$
in the Milne region is identified with period $2\pi$  by the
orbifold  action, $\theta\equiv\theta+2\pi$. Accordingly, the time
coordinate in the Rindler regions (also known as ``whiskers'')
is  also compact, $\eta\equiv\eta+2\pi$, leading to the existence of
closed time-like curves (CTC). From \eqref{xpm1},\eqref{xpm2} it
may appear that $\theta$ and $\eta$ should also be identified
modulo the complex  period $2\pi i/\beta$, as this would leave
the coordinates $x^\pm$ unchanged. There is however no sense
in identifying a real scalar field under both real and imaginary
translations, hence we only impose the real identification
$\theta\equiv\theta+2\pi$. In particular, the Rindler time
is  not  identified under imaginary translations, hence we
do not expect Misner  space to have any thermal properties.

Due to the orbifold nature of Misner space,
classical trajectories are simply straight lines on
the covering space, \be \label{clasu} X^\pm =
x_0^\pm + p^\pm \tau \ee where $M^2 = 2 p^+ p^-$. A
non-tachyonic particle with $p^+,p^->0$ therefore
comes from  the infinite past in
the Milne patch at $\tau=-\infty$  and exits  in the
future Milne patch at $\tau=+\infty$. To describe its motion along
the angular direction $\theta$, it is useful to
eliminate $\tau$ from \eqref{clasu}, and express
$\theta$ directly in terms of the cosmological time
$T$, \be \label{thetaT} \theta= \frac1{2\beta} \log
\frac{p^+}{p^-} + \frac1{2\beta} \sgn(T) \log
\frac{\sqrt{j^2+M^2 T^2}+j}{\sqrt{j^2+ M^2 T^2}-j}
\ee where $j=p^+ x_0^- - p^- x_0^+$ is the angular
momentum along the compact direction $\theta$. At
late and early times, we have
\be \theta=
\frac1{2\beta} \log  \frac{p^+}{p^-} + \sgn(T)
\left( \frac{j}{MT} - \frac{j^3}{6 (MT)^3} + \dots
\right)
\ee so that the angular velocity decreases to
zero at $T=\pm \infty$: indeed, the moment of inertia
increases to infinity as the circle decompactifies.
It is important to note
that, due to the compactness of $\theta$, $j$ is
quantized in units of $1/\beta$. Around $T\to 0$,
on the other hand, $\theta \sim \sgn(T) \log|T|$
independently of the values of $j$ and $M$: by the familiar
``spinning skater'' effect, the
particle winds faster and faster as the circle shrinks,
implying a large blue-shift factor.

For $j\neq 0$, the particle crosses
into the Rindler regions at $T=0$. There,
the expression \eqref{thetaT} still holds, upon
replacing $T \to ir$ and $\theta\to
\eta$. The resulting  particle trajectory winds
infinitely many times around
the compact time direction as it emerges from $r=0$,
propagates till
a maximum radius $r=|j|/M$, and then propagates back
towards $r=0$.
More aptly, an observer at fixed Rindler time $\eta$
will see an
infinite number of copies of the particle accumulating
towards $r=0$,
emerging from and being re-absorbed by the singularity in
a periodic
fashion. The angular momentum $j$ now describes the Rindler energy
of each of
these particles, and is quantized in accordance with the
compactness
of the time variable $\eta$. The total Rindler energy of
the configuration is thus infinite.

The infinite blue shift experienced by a particle
in the Milne patch, and the corresponding infinite
energy of the particles in the Rindler patch, have
been argued to imply, in a similar context, an
instability towards large black hole creation
\cite{Horowitz:2002mw}. Our working hypothesis
here, however, is that pair production of
twisted states may alter the geometry
sufficiently to prevent this collapse.

\subsection{Wave Functions in Misner Space}

Field theory on Misner space is a textbook example
in the study of time-dependent backgrounds (see
e.g. \cite{birrell}). Typical to  such
backgrounds it possesses different vacua, differing
due to cosmological particle production. A vacuum
of particular interest is the one inherited from
the Minkowski vacuum on the covering space:
this seems to the implicit vacuum in
first quantized string theory defined by standard
orbifold techniques.  By definition, ``positive
energy'' modes in this vacuum are superpositions of
positive energy modes on the covering space, which
are invariant under the boost identification. While
these modes may be obtained by summing over the
images of the Minkowski plane waves, it is more
useful to consider states which are eigenmodes of
the boost operator, \be \label{eig} \phi_{j,s,M^2}
(x^+, x^-) = \int_{-\infty}^{\infty} dv \exp\left[
i \left( p^+ x^- e^{-v} +  p^-  x^+  e^{v}
    \right) + i v j - v s \right]
\ee where the light-cone momenta $p^+,p^-$ satisfy the mass-shell condition $2
p^+ p^- =  M^2$, and are
positive, as appropriate for a positive
energy solution; without loss of generality, we
choose $p^+=p^- = M/\sqrt{2}$. The wave function
(\ref{eig}) of the light-cone coordinates should of
course be multiplied by a proper eigenmode in the
transverse directions, e.g. $e^{i p_i
  x^i}$, in which case the two-dimensional mass $M$ is related
to the physical mass $m$ via $M^2= m^2  +  p_i^2$.
In (\ref{eig}) we also included the dependence on
the spin $s$ in the plane $(x^+,x^-)$ (for a
complex scalar field,  an imaginary value for $s$
corresponds to the automorphic parameter in the
gravity literature). Covariance under the
boost-identification \be \phi_{j,s,M^2}
\left(e^{2\pi \beta} x^+, e^{-2\pi \beta}
x^-\right) = e^{2\pi \beta s}    \phi_{j,s,M^2}
(x^+, x^-) \ee then implies that the orbital boost
momentum $j$ should be an integer multiple of
$1/\beta$. For conciseness, we shall complexify the
orbital momentum $j$ so that $\Im(j)=s$.

As it stands, the integral in (\ref{eig}) is
ill-defined: while its oscillatory behavior yields
a finite result for spinless fields ($s=0$) in the
future Milne region ($x^+>0$ and $x^->0$), its
proper definition for all $s$ requires to deform
the $v$ integral into a contour that interpolates
between $-\infty-i\eps$ to $+\infty+i\eps$ (where
$\eps$ may be chosen to be $\pi/2$ to ensure the
best convergence.).

One  thus obtains, utilizing the standard integral
representation of the Hankel function $H^{(1)}$,
the following expression for the positive energy
wave functions in the future Milne region:
  \bea \label{han}
\phi^{+,F}_{j,s,M^2} (x^+,x^-)&=& i \pi ~ e^{\pi
j/2}~ e^{-i j \beta
  \theta}~ ~ ~
H_{-ij}^{(1)}(M T) \nn \\
&=& i \pi  ~e^{\pi j/2}~ \left(
\frac{x^+}{x^-}\right)^{-i j/2}
H_{-ij}^{(1)}\left(2 M \sqrt{x^+ x^-} \right) \eea
For $|s|<1$, it is easy to check that the integral
can in fact be performed on the real $v$-axis
without changing the result, however for general
values of the spin, it is necessary to give a small
imaginary part to $v$. This is a first departure from
standard orbifold technology.  Equivalently, one
may continue the space-time coordinates slightly
into the upper half plane, in such a way that
$\sqrt{x^+x^-}$ picks up a  small  imaginary part.

Negative energy modes on the other hand may be
obtained by choosing $p^+=p^-=-M/\sqrt{2}$. The
proper contour in \eqref{eig} then interpolates
between $-\infty+i\eps$ and $+\infty-i\eps$,
yielding the other Hankel function, \be
\label{han2} \phi^{-,F}_{j,s,M^2} (x^+,x^-) = - i
\pi  ~e^{-\pi j/2}~ \left(
\frac{x^+}{x^-}\right)^{-i j/2}
H_{-ij}^{(2)}\left(2 M \sqrt{x^+ x^-} \right) \ee
The modes \eqref{han} and \eqref{han2} form a basis
of positive and negative energy modes for the
Minkowski vacuum. From their  behavior at
$T\to+\infty$, \be \label{adia}
\phi^{\pm,F}_{j,s,M^2} (x^+,x^-) \sim e^{-i j \beta
  \theta} e^{\pm i M T}
\ee we see that they  can also be interpreted as
positive and  negative frequency modes with respect
to the adiabatic {\it out}  vacuum.

In order to define global wave functions on the
orbifold, it is necessary to analytically continue
(\ref{han})  and (\ref{han2}) to the other regions.
Since positive energy modes in Minkowski space are
analytic and bounded at the horizons in the upper
half of the complex $x^\pm$ planes, one should
analytically continue positive energy modes under $x^- \to e^{i \pi} x^-$
across the horizon at $x^+=0$, and  $x^+ \to e^{i
\pi} x^+$ across the horizon at $x^-=0$,
while negative energy modes should be continued in the
lower half complex $x^\pm$ planes. One thus
obtains modes which decay  exponentially towards
the spatial Rindler infinities, \bea \label{hanr}
\phi^{+,R}_{j,s,M^2} (x^+,x^-) &=& - i \pi ~e^{-\pi
j/2}~ \left( \frac{x^+}{- x^-}\right)^{-i j/2}
K_{-ij} \left(2 M \sqrt{(-x^-) x^+} \right) \nn\\
\phi^{+,L}_{j,s,M^2} (x^+,x^-) &=& - i \pi ~e^{-\pi
j/2}~ \left( \frac{- x^+}{ x^-}\right)^{-i j/2}
K_{-ij} \left(2 M \sqrt{(-x^+) x^-} \right) \eea
This behavior is to be contrasted
with the case of charged particles or winding
states, which, as will be recalled in Section 3,
can classically propagate to
$|r|=\infty$.

Finally, the modes \eqref{hanr} can be analytically
continued across the Rindler past horizons, by
further continuing the $\phi^R$ modes under $x^-\to
x^-  e^{i\pi}$, or equivalently the $\phi^L$  modes
under $x^+ \to x^+ e^{i\pi}$, leading to \be
\label{hanp} \phi^{+,P}_{j,s,M^2} (x^+,x^-) = i \pi
~e^{\pi j/2}~ \left( \frac{-x^+}{-x^-}\right)^{-i
j/2} H_{-ij}^{(2)}\left(2 M \sqrt{(-x^+)(-x^-)}
\right) \ee As in \eqref{adia}, it can be seen that
$\phi^{+,P}_{j,s,M^2}$ is also a positive frequency
mode with respect to the adiabatic {\it in}
vacuum: there is therefore no particle production
between the {\it in} and {\it out}  vacua,
because those are simply projections of the
Minkowski vacuum on the covering space. This is not
to say that particle production does not take place
in this cosmological background, but rather that
all particles produced at $T<0$ annihilate at $T>0$
-- a reflection of the invariance of the background
under time reversal. On the other hand, it is well
known that the ``conformal'' vacuum, defined by the
behavior of the (effectively massless) wave
functions at $T=0$, is related by a non-trivial
Bogolubov transformation to the {\it in} or {\it
out} vacua \cite{birrell}.

It is interesting to remark that the various
analytic continuations needed to define the
positive energy wave function in all quadrants can
be summarized by a simple prescription, namely
continue the space-time coordinates into the
complex upper half plane, $x^\pm \to x^\pm+i\eps$:
this amounts to inserting a regularizing factor in
\eqref{eig}, \be \label{eigr} \phi_{j,s,M^2} (x^+,
x^-) = \int_{-\infty}^{\infty} dv \exp\left[ i
\frac{M}{\sqrt{2}} \left( p^+ x^- e^{-v} +  p^- x^+
e^{v}
    \right) - \eps\sqrt{2} \cosh v + i v j - v s \right]
\ee which renders the $v$ integral convergent in
all quadrants, irrespective of the spin $s$.
Negative energy eigenfunctions with $p^+<0$ and
$p^-<0$ should instead be continued in the complex
lower half plane, $x^\pm \to x^\pm - i \eps$. This
analytic continuation of course breaks the boost
symmetry, and it is important to check that the
latter is recovered in the limit $\eps\to 0$.

Notice that this prescription fails in the case of
tachyonic wave functions with $p^+ p^- <0$;
nevertheless, it is possible to regularize
tachyonic wave functions with non zero spin $s$ by
continuing only one of the light-cone coordinates,
in a direction that depends on the sign of $s$.
This remark will become relevant below, when we
discuss long winding strings in Misner space.

Finally, let us note that wave functions may also
be regularized by constructing wave packets of
different $M^2$: while the mass-shell condition in
Minkowski space selects only one eigenvalue, it is
expected that back-reaction will deform Misner
space, in such a way that physical eigenmodes in
this geometry will be superpositions of modes with
different mass on the covering space. The resulting
modes \be \label{smo} \int \rho(M) ~ \phi_{j,s,M^2}
(x^+, x^-) dM = \int_{-\infty}^{\infty} dv
~\hat\rho\left[  \frac{1}{\sqrt{2}} \left( p^+ x^-
e^{-v} +  p^-  x^+  e^{v}
    \right)  \right]
\exp\left( + i v j - v s \right) \ee are then
unambiguously defined in all quadrants provided the
Fourier transform $\hat \rho$ of the wave profile
decreases sufficiently rapidly at $\infty$.
Implementing this within string perturbation theory
requires a better handle on back-reaction in that
framework.

\subsection{Wave Functions in Grant Space}
A close cousin of Misner space is Grant space
\cite{Grant:1992kj,Cornalba:2002fi}, i.e. the orbifold of flat
Minkowski space by the combination of a boost and a
translation,
\be \left(e^{2\pi \beta} x^+, e^{-2\pi
\beta} x^-, x+ 2\pi R \right) \equiv (x^+, x^-, x)
\ee
Since the identification has no fixed point,
the quotient is perfectly regular. Nevertheless, it
still possesses regions with Closed Timelike Curves.
Defining invariant coordinates $z^{\pm}=x^\pm
e^{\mp 2\pi \beta x/R}$, the metric can be
rewritten in a Kaluza-Klein form
\be
\label{grkk}
ds^2 = P^2
( dx + A)^2 -2 dz^+ dz^- - \frac{E^2}{P^2} (z^+
dz^- - z^- dz^+)^2
\ee where the radius $\rho$ of
the compact direction and the Kaluza-Klein electric
field $E$ are given by \be P^2 = 1 + 2 E^2 z^+
z^-\ ,\quad dA=\frac{E}{P^4}dz^+ dz^-\ ,\quad
E=\beta/R \ee
The compact direction $x$ therefore
becomes time-like  in  the region $2 z^+ z^- <
-1/E^2$, though there are  closed time-like
geodesics passing through any point with negative
$z^+ z^-$\footnote{CTCs can be eliminated by
excising the region $2 z^+ z^- < -1/E^2$, e.g. by
introducing orientifold planes
\cite{Cornalba:2002nv,Cornalba:2003ze}. We will not
discuss this construction further here.}.

Wave functions in Grant space are simply given by
the product of eigenfunctions in Misner space by
plane waves $e^{ipx}$ in the $x$ direction, upon
changing the quantization rule to $\beta j - p R
\in \Zint$. They therefore remain singular at $x^+
x^- =0$, which is now interpreted  as a chronology
horizon rather than a cosmological singularity.
Again, there is  no particle production between the
adiabatic {\it in} and {\it out} vacua. In contrast
to the Misner case however, the boost momentum $j$
can take arbitrary values (by suitably choosing
$p$), hence wave packets which are smooth across
the chronology horizon can be constructed.


\subsection{Energy-momentum Tensor in Field Theory}

As a preparation for our discussion of the 
string theory amplitude in Section 4,
we now discuss some aspects of vacuum
energy for field theory in Misner and Grant 
space \cite{Hiscock:vq,Grant:1992kj}.

Due to the existence of CTCs, both Misner space and
Grant space can be anticipated to have a divergent
one-loop energy momentum tensor (at least in field theory),
and therefore to be subject to a large
back-reaction.  The
one-loop energy-momentum tensor generated by the
quantum fluctuations of a free field $\phi$ with
(two-dimensional) mass $M^2$ and spin $s$ in Grant
space can be derived from the Feynman propagator at
coinciding points (and derivatives thereof). By
definition, any Green function in the Minkowski
vacuum is given by a sum over images of the
corresponding one on the covering space. Using a
(Lorentzian) Schwinger time representation for the
latter, we obtain \bea G(x^\mu;{x'}^\mu)=&&
\sum_{l=-\infty}^{\infty}\int_0^\infty d\rho \int
dp_\mu \exp\left[ -i p^- (x^+ - e^{2\pi\beta l}
{x'}^+) -i p^+ (x^- - e^{-2\pi\beta l}
{x'}^{-}) \right.\nn\\
&&\left. - i p_x (x-x'+2\pi R l)  + i \rho(2p^+ p^-
+ p_x^2  + p_\perp^2 + M^2)  -2\pi s l \right] \eea
where 
$s$ is the total spin carried by the field
bilinear. Performing the Gaussian integration over
the momenta $p_\mu$, one obtains \bea
\label{propco} G(x^\mu;{x'}^\mu)=&&
\sum_{l=-\infty}^{\infty}\int_0^\infty d\rho
(i\rho)^{-D/2} \exp\left[ - i(x^+ - e^{2\pi\beta l}
x^{+'})(x^- - e^{-2\pi\beta l}
x^{-'})/4\rho\right.\nn\\
&&\left. - i (x-x'+2\pi R l)^2/4\rho  + i \rho M^2
-2\pi s l \right] \eea The renormalized propagator
at coinciding points is obtained in this case by
dropping the $l=0$ contribution. As is well known,
it is related to the expectation of the energy
momentum tensor by taking two derivatives, e.g. in
the case of a massless scalar field in 4 dimensions
with general coupling $\xi$ to the Ricci curvature,
\be \label{tabxi} \langle 0 | T_{ab} | 0
\rangle_{ren}(x) = \lim_{x\to x'} \left[ (1-2\xi)
\nabla_a \nabla'_b - 2\xi \nabla_a \nabla_b +(2\xi
-\frac12) g_{ab} \nabla_c \nabla^{'c} \right]
G_{ren}(x,x') \ee which diverges like $K/T^4$ at
the cosmological singularity ($T=0$) of Misner
space \cite{Hiscock:vq,sushkov}, where the constant
$K$ is given by \be K= \sum_{n=1}^{\infty}
\cosh[2\pi n s \beta] \frac{2+\cosh 2\pi n
\beta}{(\cosh 2\pi n \beta-1)^2} \ee In the case of
Grant space, the components of the energy-momentum
tensor diverge like $1/(T^2 R^2)$ near the
chronological horizon at $z^+ z^-=0$. In
addition, there are further divergences on the
``polarized surfaces'', where the distance $R_l^2$
between one point and its $l$-th image becomes
null. This type of divergence is in fact the basis
for Hawking's chronology protection conjecture \cite{Hawking:1991nk}.
Notice that for spin $|s|>1$, the constant $K$
itself becomes infinite, a reflection of the
already noticed non-normalizability of the wave
functions for fields with spin.

In string theory, the local expectation value
$\langle 0  | T_{ab}(x) | 0 \rangle_{ren}$ is not
an on-shell quantity, hence not directly
observable. In contrast, the integrated free
energy, given by a torus amplitude, is a valid
observable. Of course, one may probe the local
structure of the one-loop energy by computing
scattering amplitudes at one-loop \cite{bdpr}. Instead,
in this Section we will compute the integrated
one-loop free energy in field theory, for the
purpose of comparison with the string theory result
in the next Section.

The integrated free energy for a spinless scalar
field of mass $M^2$ may be obtained by integrating
the propagator at coinciding points \eqref{propco}
once with respect to $M^2$, as well as over all
positions. In analogy with Section \ref{sphaleron},
it may also be computed by a path integral method,
\be Z= -\int_0^{\infty} \frac{d \rho}{\rho} \int
[DX]~e^{i S[X]} \ee where the
path integration is over all periodic paths with
period $\rho$, and the Lagrangian for a free scalar
particle in Minkowski space is \be S[X] =
\int_0^\rho\left[ 2\pa_\tau{X^+} \pa_\tau {X^-}
+(\pa_\tau{X^i})^2 \right]  d\tau + \rho M^2
\ee
Closed trajectories
in Grant space correspond to trajectories on the
covering space which close up to the action of the orbifold,
\footnote{ The integer ``twist''  $l$ along
the proper time direction should not be confused
with the winding number $w$, corresponding to the
twist in the spatial $\sigma$ direction of a closed
string.} \be X^\pm (\tau+\rho)= e^{\pm l\beta}
X^\pm(\tau) \ ,\qquad X(\tau+\rho) = X(\tau) + 2
\pi l R \ee In the semiclassical approximation, the
path integral is dominated by the classical
trajectories, which are simply straight lines on
the covering space. Their action is the square of
the geodesic distance between a point $(x^\pm,x)$
and its $l$-th image, \be R_l^2= l^2 R^2 -4
\sinh^2(l\beta/2) x^+x^-\ee
Taking into account the fluctuations around this
classical trajectory including in the $D$
transverse directions, one finds \be \label{ftg}
\mathcal{F} = \sum_{l=-\infty}^{\infty} \int dx^+
dx^- \int_0^\infty
\frac{d\rho}{(i\rho)^{\frac{D}{2}-1}} \exp\left( -8
i \sinh^2 (\pi \beta l) x^+ x^- -i (2\pi R l)^2  +
i M^2 \rho  \right) \ee In contrast to the flat
space case, the integral over the zero-modes
$x^\pm,x$ does not reduce to a volume factor, but
gives a Gaussian integral, centered on the light
cone $x^+ x^-=0$. Dropping as usual the divergent
$l=0$ flat-space contribution, one obtains a finite
result \be \label{ft} \mathcal{F} =
\sum_{l=-\infty, l\neq 0}^{+\infty} \int_0^{\infty}
\frac{d\rho}{\rho^{1-\frac{D}{2}} }\frac{e^{- M^2
\rho - \frac{\pi^2 R^2 l^2}{\rho_2} -2\pi \beta s
l} } {\sinh^2\left( \pi \beta l\right)} = i \ln
\langle adia, {\it out} | adia, {\it in} \rangle
\ee where we reinstated the dependence on the spin
$s$ of the particle. As written out  on the
right-hand side, the integrated free energy should
be interpreted as the logarithm of the transition
amplitude between the adiabatic {\it in} and {\it
out}  vacua. Consistently with our analysis  of the
positive energy modes in Section 2.2, ${\cal F}$ does not have any
imaginary part, implying  the absence of net
particle production between  past and future
infinity. Additionally the free energy is infrared
finite for each spin $s$, even for the Misner
($R=0$) limit. We shall return to \eqref{ft} in Section 4.2,
to compare it with the string theory one-loop amplitude
in the untwisted sector.

\section{Twisted Strings in Misner Space}

In this Section, we clarify the nature of physical
twisted states, pursuing the discussion in
\cite{Pioline:2003bs}. One of the main insights,
based on the analogy with the charged open string
in an electric field, was that care should be
exercised in quantizing the bosonic zero-modes : in
contrast to the Euclidean rotation orbifold, where
the string center of mass is effectively confined
by an harmonic potential, in the Lorentzian
orbifold case the center of mass moves in an  {\it
inverted} harmonic potential, hence has a
continuous spectrum. On the other hand, for excited
modes, it was found that the standard Fock space
quantization, valid in flat space, was also
appropriate in the Lorentzian orbifold. While
fermionic modes were not discussed in
\cite{Pioline:2003bs}, it is easy to see that
both zero-modes and excited modes
should be  treated in the same way as in flat
space: indeed, they lead to imaginary values for
$L_0$ and $\tilde L_0$, as appropriate for states
carrying a non-zero spin\footnote{ See the
analogous statement for longitudinal gauge bosons,
\cite{Pioline:2003bs} eq. (3.46).}. We thus
restrict our attention to the bosonic zero-modes,
which are responsible for the non-trivial
space-time dependence of the wave functions.

In Section 3.1 we recall the basic first
quantization formulae for closed strings in the
Lorentzian orbifold, following the discussion of
\cite{Pioline:2003bs}. In Section 3.2, we discuss
the classical string trajectories, and find that
physical states in the twisted sectors come in two
kinds: {\it short strings}, which wind along the
compact space-like direction in the cosmological
(Milne) region, and {\it long strings}, which wind
along the compact time direction in the (Rindler)
whiskers. The latter can be viewed as infinitely
long static open strings, stretching from Rindler
infinity to a finite radius and folding back onto
themselves.

We then turn to the problem of first quantization,
and we find two convenient representations of
spacetime modes of the twisted sector fields. In
Section 3.3 we use the relation of twisted closed
strings to  a charged particle in an electric
field, clarifying the analogy used in
\cite{Pioline:2003bs}. For this set of modes, just
as in the Schwinger effect, we find that local
production of pairs of long or short strings occurs
as a result of tunneling under the potential
barrier in the Rindler patches. An alternative
representation of the zero-mode algebra, which we
call the oscillator representation, is introduced
in Section 3.4, and the relation between both sets
of modes is described there.

Finally, we take some preliminary steps towards   second
quantization   in Section 3.5, concentrating mainly
on the vacuum structure of the space. We find that
in addition to the familiar vacuum ambiguity,
string theory   exhibits more options, which we
call "charged particle vacua", as they utilize the
analogy with the problem of charged particle in an
electric field. We comment on the advantages and
drawbacks of the suggested quantization schemes.
Irrespective of the choice of vacua, the general
conclusion is that Bogolubov coefficients are given by the
tree-level two-point function of twisted fields,
after properly identifying the basis of {\it in}
and {\it out } states.
The pair production rate thus follows from the choice of basis
of the zero-mode wave functions only.

\subsection{Review of First Quantization}
Using the same conventions as in
\cite{Pioline:2003bs}, we recall that the string
quasi-zero mode (thus named because its frequency
can be much below the string scale) takes the
simple form \be \label{x0} X_0^\pm(\tau,\sigma)=
\pm \frac1{2\nu} \alpha_0^\pm e^{\pm
\nu(\tau-\sigma)} \mp \frac1{2\nu} \talpha_0^\pm
e^{\mp \nu(\tau+\sigma)} \ee where $\nu=-\beta
w$ is the product of the boost parameter $\beta$ of
the orbifold, and the winding number $w$. An
important feature of the solution \eqref{x0} is
that it depends on the world-sheet coordinate
$\sigma$ only by a boost,
$X_0^\pm(\sigma,\tau)=e^{\mp\nu\sigma}
X_0^\pm(0,\tau)$, ensuring the appropriate
periodicity condition $X^{\pm}(\sigma+2\pi,\tau)
=e^{\pm w \beta} X^{\pm}(\sigma,\tau)$. At a given
time $\tau$, the world-sheet therefore wraps $w$
times the compact Milne or Rindler circle. The
radial coordinate on the other hand depends on
$\tau$ as \be \label{xpxm} 2 \nu^2 (X^+ X^-+ X^-
X^+) = \alpha_0^+ \talpha_0^- e^{2\nu\tau} +
\alpha_0^- \talpha_0^+ e^{-2\nu\tau} -  \frac12
\left( M^2 + \tM^2 \right) \ee Here we used the
Virasoro conditions, which require \be \label{vir}
M^2 = \alpha_0^+ \alpha_0^- + \alpha_0^- \alpha_0^+
\ ,\quad \tilde M^2 = \talpha_0^+ \talpha_0^-  +
\talpha_0^- \talpha_0^+ \ee where $M^2$ and $\tilde
M^2$ depend on the oscillator numbers for the
excited modes, and the momentum of the remaining
spatial directions, e.g. in the bosonic string
case, \bea \label{mm} M^2 &=& \nu^2   - 2
\sum_{m=1}^{\infty} \left( \a_{-m}^+ \a_m^-  +
\a_{-m}^- \a_{m}^+ \right) + N_L - \frac{1}{12}
\nn\\
\tM^2 &=& \nu^2   - 2 \sum_{m=1}^{\infty} \left(
\ta_{-m}^+ \ta_m^-  + \ta_{-m}^- \ta_{m}^+ \right)
+ \tilde N_R - \frac{1}{12} \eea where $N_L$ and
$\tilde N_R$ are the left-moving and right-moving
contributions from the conformal field theory
describing the remaining transverse
directions\footnote{For the superstring, fermionic
oscillators would contribute to $N_L$ and $N_R$;
the vacuum energy $-\nu ^2$ in the Neveu-Schwarz
sector cancels the first term in \eqref{mm}, while
the vacuum energy $-i\nu(1-i\nu)$ in the Ramond
sector remains leaves an non-zero imaginary part, in accordance with the
fact that the Ramond ground state carries
half-integer spin.}. Canonical quantization imposes
the commutation relations \bea [\alpha_m^+,
\alpha_n^-] = -(m+i\nu) \delta_{m+n}\ &,&\quad
[\tilde \alpha_m^+, \tilde \alpha_n^-] = -(m-i\nu)
\delta_{m+n}\ ,\quad \label{cancon}
\nn\\
(\alpha_{-n}^\pm)^* = \alpha_n^\pm\ &,&\quad
(\tilde \alpha_{-n}^\pm)^* = \tilde \alpha_n^\pm
\eea
In particular, the zero-mode algebra will be of particular interest:
\be
\label{0m}
[\a_0^+, \a_0^-]=-i\nu\ ,
\quad[\talpha_0^+, \talpha_0^-]=i\nu
\ee
Notice from \eqref{cancon} that
a creation oscillator $\a^\pm_{-n}$
contributes an imaginary part $\pm i\nu$ to $M^2$,
while $\ta^\pm_{-n}$ contributes an imaginary part
$\mp i\nu$ to $\tM^2$. Denoting \bea
h&=&\sum_{m=1}^{\infty} \left( \frac{\alpha_{-m}^-
\alpha_{m}^+ }{m+i\nu}
-\frac{\alpha_{-m}^+  \alpha_{m}^- }{m-i\nu} \right)
\in\Zint \nn \\
\tilde h&=&\sum_{m=1}^{\infty} \left(
\frac{\talpha_{-m}^-  \talpha_{m}^+   }{m-i\nu} -
\frac{\talpha_{-m}^+  \talpha_{m}^- }{m+i\nu}
 \right) \in \Zint
\eea (plus the integer or half-integer valued
fermionic contributions in the Neveu-Schwarz or
Ramond sectors of the superstring) we see that the
zero-mode boost operator \be \label{jzero} j=
\frac{1}{2\nu} \left(
 \alpha^+_0\alpha^-_0 +
\alpha^+_0\alpha^-_0 \right) -\frac{1}{2\nu} \left(
\talpha^+_0\talpha^-_0 + \talpha^-_0\talpha^+_0
\right) = \frac{1}{2\nu} \left(M^2 - \tM^2\right)
\ee picks up an imaginary part $\Im(j)=h + \tilde
h$ equal to the spin of the state in the
$(x^+,x^-)$ plane, in accordance with our
discussion in Section 2. In addition, the
total mass $\mu^2=(M^2+\tM^2)/2$ receives an non-zero
imaginary part $2\nu i (h-\tilde h)$ for distinct
left and right-moving helicities. We shall be
interested in non tachyonic states only, such that
$\Re(M^2)\geq 0, \Re(\tM^2)\geq 0$.

\subsection{Semiclassical Analysis: Short and Long Strings}

At the classical level, we may choose the origin of
$\tau$ and $\sigma$ such that $|\a_0^+| = |\a_0^-|$
and $|\ta_0^+| = |\ta_0^-|$. We are thus left with
two choices of sign,
\be \a_0^+ = \a_0^- = \eps
M/\sqrt{2}\ ,\quad \ta_0^+ = \ta_0^- = \tilde \eps
\tM/\sqrt{2}
\ ,\quad \eps=\pm 1\ , \tilde\eps=\pm 1
\ee
As apparent from \eqref{xpxm}, the
location of the string at  large world-sheet time
$|\tau|$ depends crucially on the sign of the
product: for $\eps \tilde\eps>1$, the string starts
and ends at large positive $X^+ X^-$, and therefore
in the Milne region; in contrast, for  $\eps
\tilde\eps<1$, it start and ends in the Rindler
regions. Restricting to $j=0$ for simplicity, the
zero-more indeed takes one of two simple forms,
\begin{itemize}
\item $(\eps,\teps)=(1,1)$ {\it (short string)}:
\be X^\pm(\sigma,\tau) = \frac{M}{\nu\sqrt{2}}
\sinh(\nu  \tau) e^{\pm \nu \sigma} \ ,\quad
T=\frac{M}{\nu}\sinh(\nu  \tau)\ ,\quad \theta=w
\sigma \ee describes a string winding around the
Milne circle, and propagating from past infinity
$T=-\infty$ to future infinity $T=+\infty$ as
world-sheet time $\tau$ passes. The opposite case
$(\eps,\tilde\eps)=(-1,-1)$ is just the time
reversal of this process.
\item $(\eps,\tilde\eps)=(1,-1)$ {\it (long string)}:
\be X^\pm(\sigma,\tau) = \pm \frac{M}{\nu\sqrt{2}}
\cosh(\nu  \tau) e^{\pm \nu \sigma} \ ,\quad
r=\frac{M}{\nu}\cosh(\nu  \tau)\ ,\quad \eta=w
\sigma \ee in contrast is entirely contained in the
right Rindler patch: from the point of view
of an observer at fixed Rindler time, it can be viewed as a static string
stretching from Rindler infinity  $r=\infty$ to a
finite radius $r_0=M/\nu$, folding onto
itself and extending back to $r=\infty$ again.  The opposite choice
$(\eps,\tilde\eps)=(-1,1)$ describes a stretched
static string in the left Rindler patch. Each of
these configurations is invariant under time reversal.
\end{itemize}
\FIGURE{ \label{shortlongfig}
\hfill\epsfig{file=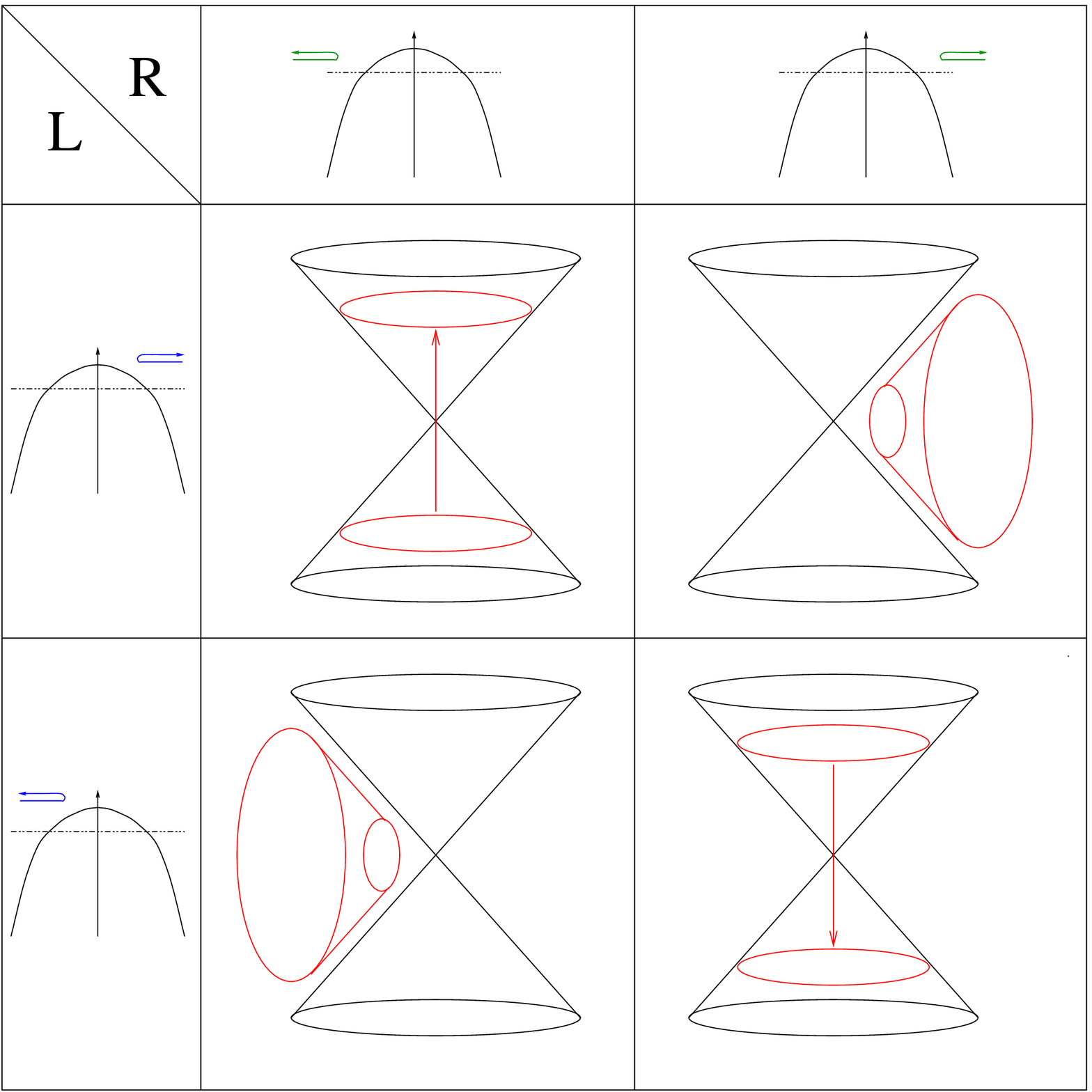,height=9cm}
\hfill \caption{Winding strings on the Lorentzian
orbifold exist in two different species: short
strings (wrapping the Milne circle) and long
strings (stretched from Rindler infinity to a
finite radius and folded back onto themselves). The
inverted harmonic potentials refer to the
symmetric quantization scheme of Sec. (3.4).}} In view of the
very different nature of these  states, we shall
refer to them as {\it short} and {\it long} winding
strings, respectively. For non-vanishing $j$, the
situation is much the same, except that the
extremal radius reached by the string is now \be
\label{tpt} r_0^2 = -2 X^+ X^- =
\frac{1}{4\nu^2}  (M -\eps\tilde\eps \tilde M) ^2
\ee Long strings therefore stay away from the
origin of Rindler patch, while short strings wander
for a finite interval $[\tau_0,\tau_1]$ into one of
the Rindler patches, depending on the sign of $j$:
again, in this region they can be viewed as static
strings, which extend from the origin of Rindler
space to a finite radius, and fold back onto
themselves. In contrast to the long strings, they
have finite length at every finite cosmological time $T$.
The fact that a string
world-sheet winding around a compact time direction
is better thought of as a static string is quite
general.

It is important to note that the induced metric on
the long string solution has unusual causality
properties: the world-sheet $\sigma$ direction wraps
the target-space time-like direction $\eta$, while
the world-sheet $\tau$ direction generates a motion
in the space-like radial direction $r$. This also
takes place during the interval $[\tau_0,\tau_1]$
for short strings with $j\neq 0$, where the induced
metric undergoes a signature flip as the string
crosses the horizon. Such a behavior is reminiscent
to the case of classical trajectories for tachyonic
states in flat space, however here it is found for
otherwise sensible physical states with $M^2>0,
\tM^2>0$. More aptly, it is analogous to the case
of supertubes or long strings in G\"odel
Universe\cite{Drukker:2003sc, Israel:2003cx}).

There seems to be however no reason to exclude
these solutions a priori, since they are bona fide
solutions of the Polyakov action in the conformal
gauge \footnote{Recall that the equations of motion
of the Polyakov action imply the equality of the
induced and world-sheet metric only up to a
conformal factor, which can be used to flip the
signature of the world-sheet.}. We will return to
this point in Section \ref{rindlerq}, where we
quantize the strings from the point of view of an
observer in the Rindler patches.

Finally, the above classical solutions carry over to Grant
space, after modifying the quantization condition on $j$
appropriately. The single-valued coordinates
$z^\pm$ introduced in \eqref{grkk} are now independent
of the $\sigma$ coordinate,
\be
\label{x0gr} Z_0^\pm(\tau,\sigma)=
\pm \frac1{2\nu} \alpha_0^\pm e^{\pm
\nu\tau} \mp \frac1{2\nu} \talpha_0^\pm
e^{\mp \nu\tau} \ee
so that the string center of motion follows a particular world-line in the
Grant geometry, while the string itself wraps the compact coordinate $x$.
It is easy to see that the matching conditions imply that, for long
strings, the extremal radius $r_0$ always lies in the region where
the coordinate $x$ is time-like. Short strings on the other hand
come from infinite past, cross the chronological horizon $z^+z^-=0$
and return into the future patch, much as in the original Misner
geometry. For appropriate choices of $j$ and $p$, they may cross
over into the $r_0>1/E$ region.

\subsection{From Twisted Closed Strings to Charged Particles \label{opcl}}
As observed in \cite{Pioline:2003bs}, the closed
string zero-mode \eqref{x0} is strikingly
reminiscent of the classical trajectory of a
charged particle (or string) in an electric field
\be X^\pm(\tau) = x_0^\pm \pm \frac{a_0^\pm}{\nu}
e^{\pm \nu \tau} \ee Clearly, the world-line of the
charged particle mode $(x_0^\pm, a_0^\pm)$ can be
mapped to the motion of a point at fixed
$\tau+\sigma$ on the twisted closed string
world-sheet, upon identifying \be \label{idop}
x_0^\pm = \mp \frac{1}{2\nu} \ta_0^\pm\ ,\quad
a_0^\pm = \frac{1}{2} \a_0^\pm \ee Since the closed
string zero-mode \eqref{x0} depends on $\sigma$
only through the boost action, it is thus
straightforward to obtain the entire closed string
world-sheet by smearing the charged particle
trajectories under the boost action. Charged
particle trajectories which remain inside the
Rindler patches are thus associated to long string
configurations, while charged particle trajectories
which cross the Rindler horizon are associated to
short strings.

\FIGURE{ \label{cutting}
\epsfig{file=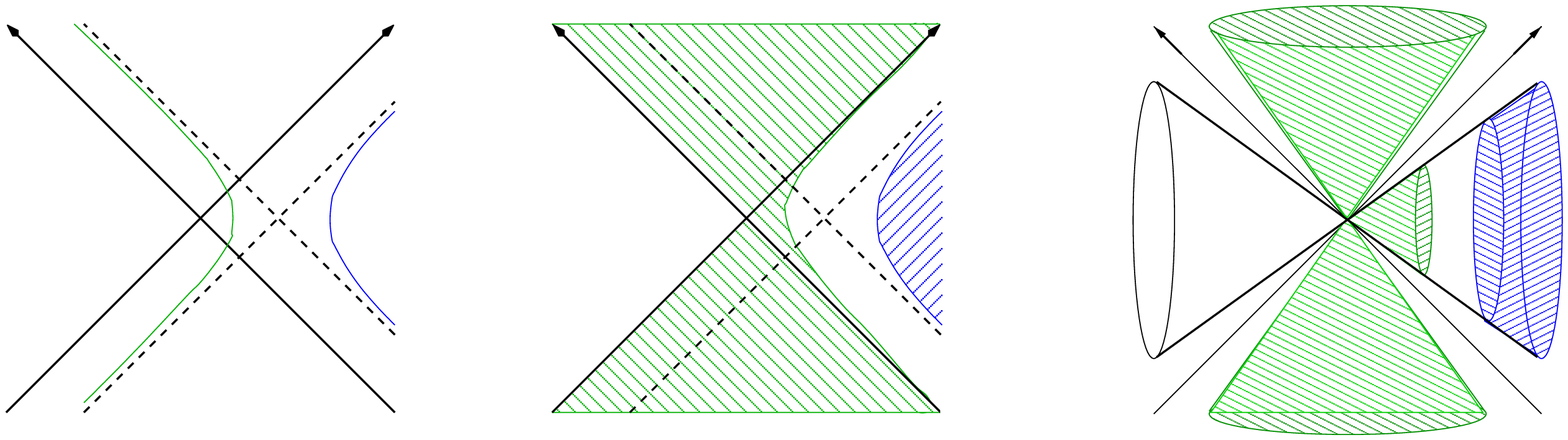,height=4cm}
\caption{
The twisted string world-sheet can be obtained by smearing the 
charged particle trajectory under the action of the boost.
Charged particles crossing the Rindler horizon correspond to
short strings (in green), while charged particles escaping
to Rindler infinity correspond to long strings (in blue).}
}

This embedding can be used to simplify the first
quantization of the twisted closed string. At the
quantum level, the closed string zero-modes algebra
\eqref{0m} is also isomorphic to the charged
particle algebra under the identification
\eqref{idop}. This implies that we can use the same
representation in terms of covariant derivatives
acting on wave functions of two coordinates
$x_+,x_-$ as in the charged particle case, \be
\label{reals} \a_0^\pm= i \nabla_{\mp} = i \p_\mp
\pm \frac{\nu}{2} x^\pm\ ,\quad \talpha_0^\pm= i
\tilde \nabla_{\mp} = i \p_\mp \mp \frac{\nu}{2}
x^\pm \ee in such a way that the physical state
conditions are simply the Klein-Gordon operators
for a particle with charge $\pm \nu$ in a constant
electric field, \be M^2= \nabla_+ \nabla_- +
\nabla_- \nabla_+\ ,\quad \tM^2= \tilde\nabla_+
\tilde\nabla_- + \tilde\nabla_- \tilde\nabla_+ \ee
To understand the physical meaning of the $x^\pm$
coordinates in the closed string setting, observe
that in this representation, the zero-mode
\eqref{x0} reads \be X^{\pm}_0(\sigma,\tau) =
e^{\mp \nu \sigma} \left[ \frac12 \cosh (\nu \tau)
~  x^\pm \pm \frac{i}{\nu} \sinh (\nu \tau) ~
\pa_\mp \right] = \frac12 x^\pm
~~~\mbox{at}~\tau=\sigma=0 \ee The coordinates
$x^\pm$ are thus (up to a factor of 2) the
(Schr\"odinger picture) operators corresponding to
the location of the closed string at $\sigma=0$.
The radial coordinate $r=\sqrt{-2 x^+x^-}$ associated
to the
coordinate representation \eqref{reals}, should be
thought of as the radial position of the closed
string in the Rindler wedges
at a fixed $\tau$. On the other hand
 the world-sheet   dependence on the angular coordinate (for a
given  winding number)
  is fixed in terms of the boost
momentum $j$.

The physical state conditions are most usefully
imposed in this representation on $\mu^2=(M^2+\tM^2)/2$ and on
$j=(M^2-\tilde{M}^2)/(2\nu)$.  The wave
functions for the radial location of the twisted
closed string modes are thus the same as the radial
part of the quantum wave functions for charged
particles in an electric field, with boost momentum
$j=i (x^+ \pa_+ - x^- \pa_-)$ given by the
quantized value \eqref{jzero}.

We can therefore make use of the  discussion of
charged particle wave functions in Rindler
coordinates \cite{Pioline:2003bs,Gabriel:1999yz},
and obtain the wave functions in the right Rindler
patch as zero-energy modes of the Schr\"odinger
operator $- \frac{d^2}{dy^2} + V(y) = 0$, with
potential \be \label{schror} V(y) = M^2 r^2 -
\left( j + \frac12 \nu r^2 \right)^2 = \frac{M^2
\tM^2}{\nu ^2} - \left( \frac{M^2+\tM^2}{2\nu} -
\frac{\nu}{2} r^2 \right)^2 \ee where $r=e^y$.
Assuming that the state is non-tachyonic, i.e. $M^2>0$
and $\tM^2>0$, the potential includes a classically
forbidden region, with turning points given by
\eqref{tpt}. Depending on the sign of $j$, this
corresponds to the tunneling of a long string
coming from Rindler infinity into a short string
going forward ($j<0$) or backward ($j>0$) in time.
The wave functions ${\cal U}_{in,R}^j$ and ${\cal
V}_{in,R}^j$ introduced in \cite{Gabriel:1999yz,Pioline:2003bs}
can now be interpreted as operators creating a long string
stretching from infinity, resp. a short string
stretching into the horizon. 

It is perhaps enlightening to describe the tunneling process
semi-classically. As usual in ordinary quantum mechanics, tunneling
can be described by continuing the proper time
$\tau$ to the imaginary axis, i.e. solving the Schr\"odinger
equation \eqref{schror} in the flipped potential $-V(y)$.
One thus obtains an Euclidean world-sheet stretching between the
long and short string world-sheet, i.e. between two hyperbolas
centered at the origin in the Rindler wedges. Alternatively, one may keep
using a Lorentzian world-sheet but analytically continue the target
space, i.e. rotate $\nu\to i\nu$ as well as change the reality
conditions on $X^\pm$ to $(X^\pm)^*=\pm X^\mp$. This maps 
the Lorentzian orbifold in the Rindler regions\footnote{Continuing the
Milne regions in this fashion would give rise to two times.}
to the Euclidean rotation orbifold, $Z \equiv e^{i\beta} Z$.
The tunneling trajectories now describe closed strings trapped
at the tip of the conical singularity and oscillating between
an inner and outer radius $r^\pm$. It would be tempting to use this
analytic continuation at $X^0<0$ to define a Hartle-Hawking type of 
vacuum in the Rindler regions, however such a prescription 
does not seem to be consistent with the orbifold identification. 
We shall return to this issue in Section 4, and argue that the
correct picture to describe winding string production is the
``EM'' one: Euclidean world-sheet and Lorentzian target space.

For a given choice of vacuum,
we can now obtain the pair production rate by simply evaluating
the Bogolubov coefficients, i.e. the overlap of the
corresponding wave functions. The transmission coefficients
in the Rindler and Milne regions can be read off 
from \cite{Gabriel:1999yz,Pioline:2003bs},
\be
\label{ref}
q_2 =
 e^{-\pi M^2/2\nu} 
\frac{|\sinh \pi j|}{\cosh\left[ \pi \tM^2/ 2\nu\right]}\ ,\quad
q_4 =
 e^{-\pi M^2/2\nu} 
\frac{\cosh\left[ \pi \tM^2/ 2\nu\right]}{|\sinh \pi j|}\ ,
\ee
respectively.
As in the Schwinger effect, tunneling corresponds
to induced pair production, and implies that
spontaneous pair production takes place as well.
{}From \eqref{ref}, one notices that the production rate
in the Milne regions $q_4$ is infinite for vanishing boost momentum $j=0$,
while the production rate in the Rindler regions vanishes. 
This can
be traced to the fact that the energy of the tunneling process
becomes equal to the potential energy at the horizon $y\to \infty$,
leading to the production of a non-normalizable state. 
For $j$ non-zero, and holding the contribution of excited and
transverse modes $M^2,\tM^2$ fixed, the rate $q_4$ diverges
linearly in $w$ as the winding number grows to infinity\footnote{In
 this discussion, we disregard the fact that $j$ is quantized
in multiples of $1/beta$.}.
As we shall see in Section 5, this can be viewed as
a consequence of the singular geometry, and resolving
it leads to a finite production rate at $j=0$ (or $w=\infty$).

Given the wave function in the Rindler patch, there
are many ways to analytically continue it across
the horizons to other patches, consistently with
charge conservation. Among these various choices,
those which involve only analytic continuation into
the upper half of the complex $x^+$ and $x^-$
planes play a special r\^ole: they have support on
positive energy states as measured by a
static observer in the Minkowski covering
space\footnote{As apparent from \eqref{schror}, the
behavior towards the horizon $y\to-\infty$ does not
depend on the mass nor on the charge, but only on
the boost momentum $j$. The usual bounded
analyticity statement about positive energy modes
of a massless scalar field therefore extends to our
case as well.}. These global modes are usually
known as Unruh modes in the neutral case, and we
will keep this name in our case as well. In Figure
\ref{unruhfig}, we have represented two examples of
Unruh modes for a charged particle, and their
interpretation in terms of winding strings. Just as
in the untwisted sector case, Unruh modes should be
the positive energy modes implicit in the
perturbative string expansion.

\FIGURE{ \label{unruhfig}
\epsfig{file=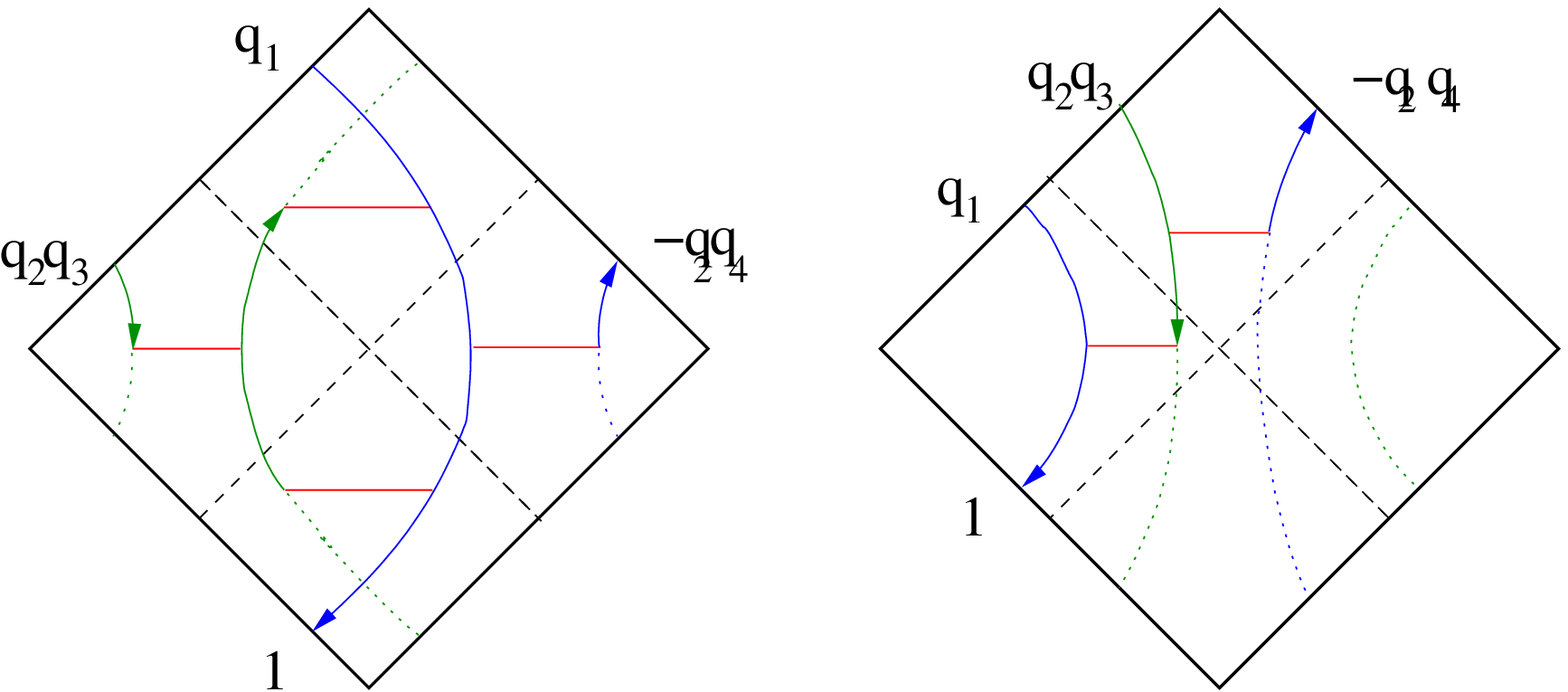,height=3cm} \hfill
\epsfig{file=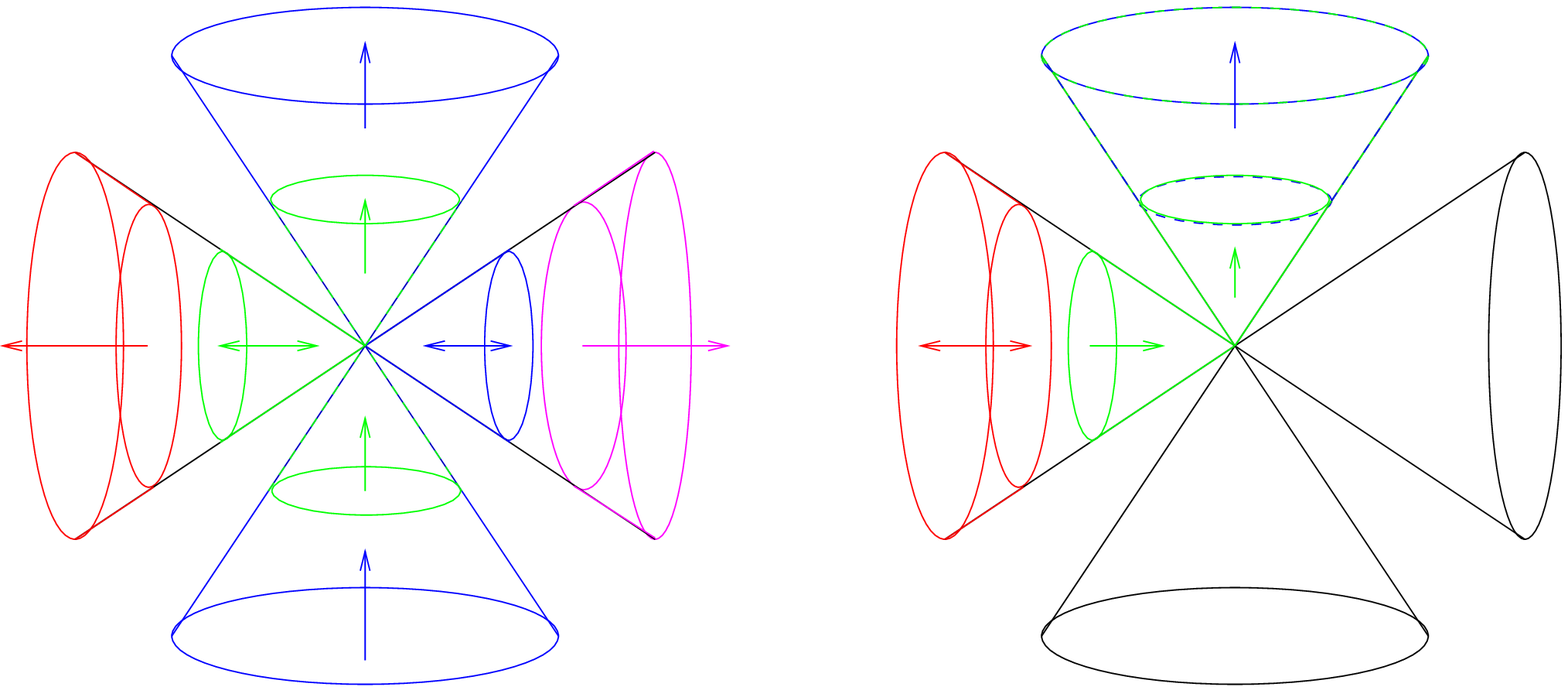,height=3cm}
\caption{Global Unruh modes for a charged particle
in an electric field (left), and their
interpretation as closed string wave functions
(right). The latter are obtained by smearing the
charged particle wave function over the action of
continuous boosts.}}

\subsection{Oscillator Representation}
We now come to an alternative representation of the
closed string zero modes, which treats left and
right movers symmetrically. The zero mode algebra
\eqref{0m} consists of two harmonic oscillators.
Motivated by the standard treatment of the harmonic
oscillator we can represent the zero mode on the
space of functions $f(  \a^+,  \talpha^- )$, with
\be \a^-=i\nu\partial_{\a^+},\ \ \
\talpha^+=i\nu\partial_{\talpha^-} \label{oscr} \ee
The Virasoro constraints  then requires that the
wave function depends on $\a^+,  \talpha^-$ as a
power, \be \label{eigo} f( \a^+,  \talpha^- ) =
\left( \eps ~ \a^+ \right)^{\frac{M^2}{2i\nu} -
\frac12} \left( \teps~ \ta^-
\right)^{\frac{\tM^2}{2i\nu} - \frac12}\ , \ee
up to a normalization factor. The
choice of signs $\eps,\teps$  is correlated with
the short and  long string branches discussed in
Section 3.2. In addition to its simplicity, the
interest of this representation stems from the fact
that eigenstates \eqref{eigo} can be obtained from
the oscillator states $(a^\dagger)^n (\tilde
a^\dagger)^{\tilde  n}$ of the standard (Euclidean)
harmonic oscillator by analytically continuing the
integers $(n,\bar n)$ to complex values
\cite{bdpr}.

The relation of this representation to the
real-space $(x^+,x^-)$ representation \eqref{reals}
may be obtained by diagonalizing simultaneously
$\a^+$ and $\ta^-$, leading to \be f(x^+,x^-) =
\int d\alpha^+ d\talpha^- \Phi^+_{ \nu, \a^+,
\talpha^-}(x^+,x^-) f(  \a^+,  \talpha^-) \ee where
\be \label{phip} \Phi^+_{ \nu, \a^+,
\talpha^-}(x^+,x^-) = \exp\left( \frac{i\nu x^+
x^-}{2} - i \a^+ x^- -i \talpha^- x^+ +
\frac{i}{\nu} \a^+ \talpha^- \right) \ee
Alternatively, one may choose to diagonalize $\a^-$
and $\ta^+$, leading to \be \label{intw} f(x^+,x^-)
= \int d\talpha^+ d\alpha^-~ \Phi^+_{ \nu, \ta^+,
\alpha^-}(x^+,x^-)~ f(  \ta^+, \alpha^-) \ee where
\be \label{phim} \Phi^-_{ \nu, \talpha^+,
\alpha^-}(x^+,x^-) = \exp\left( - \frac{i\nu x^+
x^-}{2} - i \ta^+ x^- -i \alpha^- x^+ -
\frac{i}{\nu} \ta^+ \alpha^- \right) \ee 
On shell wave functions in this representation are 
related to \eqref{eigo} by Fourier transform, and 
are now given by
\be \label{eigo2} f( \ta^+,  \alpha^- ) =
\left( \teps ~ \ta^+ \right)^{-\frac{\tM^2}{2i\nu} -
\frac12} \left( \eps~ \a^-
\right)^{-\frac{M^2}{2i\nu} - \frac12} \ee
up to normalization.
The kernels \eqref{phip} and \eqref{phim} can be
thought as the analogue, for  twisted states, of
the off-shell wave functions $e^{-i k^+ x^- -i k^-
x^+}$  in Minkowski space.

Upon substituting in \eqref{intw} the wave function
\eqref{eigo}, and integrating over $\a^+$, $\ta^-$
in one of the four quadrants, one obtains in
principle an on-shell wave function -- provided the
integral is well defined. Defining $\a^+ = \eps u
e^v, \ta^- = \teps u e^{-v}$ with $u>0$, one can
first perform the angular integral, obtaining the
same wave functions as in the untwisted case: for a
short string moving forward in time
($\eps=\teps=1$), one obtains \be f(x^+,x^-) =
\int_{0}^{\infty} dr \left[ \left(\frac{x^+}{x^-}
\right)^{-ij/2} H^{(1)}_{-ij} (2 r \sqrt{x^+ x^-})
\right] r^{\frac{i(M^2+\tilde M^2)}{2\nu}} e^{i
r^2/\nu} \ee with the Hankel function $H^{(1)}$
being replaced by $H^{(2)}$ in the time-reversed
case. On the other hand, for a long string in the
right Rindler patch ($\eps=-\teps=-1$), the
wave function appearing after
the $v$ integration corresponds to the wave function
for a tachyonic untwisted state.
Again, we stress that long strings are nevertheless
non-tachyonic physical states with $M^2,\tM^2>0$.

It is also instructive to carry out the integration in reversed order.
The radial integration leads to a parabolic cylinder function,
which in the limit of large $|v|$ is dominated by the saddle point
value,
\be
f(x^+,x^-) \propto
\int_{-\infty}^{\infty} dv
\left[ 
\left(
 \teps x^- e^{v} + \eps x^+ e^{-v} \right)
\right]^{\frac{M^2+\tM^2}{2i\nu}}
\exp\left[-\frac{i\nu \eps \teps}{4} \left( (x^+)^2
e^{-2v} + (x^-)^2 e^{2v} \right) -i j v\right] \ee
Comparing to the untwisted case \eqref{eig}, we see
that this expression is well defined as an integral
over the real $v$ axis only when the spin
$s=\Im(j)<2$, which is a (minor) improvement over
the untwisted case. For higher values one should
deform the $v$ integration contour in the complex
plane as in \eqref{eig}. Equivalently, one should
continue $x^\pm \to x^\pm -i~\sgn(\eps\teps  x^\pm) ~0^+$:
the required analytic continuation is thus different
for long and short strings.
We shall return to the  analyticity properties of
these wave functions when we discuss second
quantization  in Section 3.5.1 below.

\subsection{Second Quantization}

Having described the first quantization of the
twisted sector zero-modes $\a_0^\pm,\ta_0^\pm$ on
the closed string world-sheet, we now come to the
issue of the second quantization of the resulting
continuum of physical   states. As we are dealing
with a time-dependent geometry, we expect a variety
of vacua, depending on the kind of observer:
\begin{itemize}
\item[(i)]
String perturbation theory presumably picks the ``Minkowski'' vacuum,
defined with respect to the $x^0$ time variable on the covering space.
While creation and annihilation operators are usually
discriminated by their analyticity properties in
the complex $x^0$ plane, we will see that long
strings have no such analyticity property (in line
with their other
 tachyon-like
features) and an ad hoc prescription must be devised for
them: we
propose that they be distinguished by the sign of their
  boost momentum $j$ , which generates time evolution in the
Rindler patches. We will show in Section 4.2 that the
string vacuum amplitude has no imaginary part, which implies
that there should be no overall pair production in this vacuum.

\item[(ii)] On the other hand, one may be interested
in the vacuum defined by an accelerated observer in
the Rindler patch, i.e. an observer sitting at a
fixed radius in one of the Rindler patches. This
choice is particularly appealing since the geometry
in the Rindler patch is time-independent, and
states can be quantized unambiguously according to
the sign of their energy (after deciding on an
arrow of time in each of the Rindler regions). In
addition, the line $x^0=0$ is the most natural
Cauchy surface in the geometry\footnote{It is not
strictly speaking a Cauchy surface due to the
identification of Rindler time $\eta$, however this
identification can be imposed by restricting to
states with quantized Rindler energy.}, and one may
describe the global quantum state as a state in the
tensor product of the left and right Rindler
patches. This would allow a description of the
cosmological evolution in the Milne patch by
standard time-independent techniques, adapted to
the existence of CTCs. As we shall see, one problem
with this vacuum is that short strings have an
unbounded negative energy spectrum, which, at least
in the absence of CTCs, would lead to an
instability.
\item[(iii)] As in the untwisted case, one may choose to define
a ``conformal'' vacuum in terms of the behavior as $T\to 0$ in the
Milne patch. This procedure will however miss the long string states,
which live in the Rindler patches only and never approach the horizon.
We will not discuss these vacua any further here.
\item[(iv)] More radically, one may choose to quantize with respect to the
radial coordinate $r$ in the Rindler patches. This
procedure would be natural from an holographic
point of view, where one would impose boundary
conditions at Rindler infinity, and let them evolve
into the bulk. This is already suggested by the
long strings, which propagate radially rather than
forward in time under the world-sheet time
evolution. However, this procedure would miss the
short string states, which do not propagate to
Rindler infinity.
\end{itemize}
Finally, one may push the analogy to the charged
particle  to the level of second quantization, and
devise a closed string vacuum modeled on the $in$
and $out$ vacua for charged particles in an
electric field. Since treating the left and right
moving zero-modes $\a_0^\pm$ and $\ta_0^\pm$
symmetrically and independently does not   lead to
an unambiguous prescription for the long strings,
one may instead choose to identify one side with
the center of motion coordinate $x_0^\pm$ and the
other with the velocity $a_0^\pm$ of the charged
particle (as in Section 3.3), and inherit a well
defined prescription for $in$ and $out$ vacua.
However, making the opposite choice would lead to
different vacua, so the physical interpretation  of
these vacua is unclear. This is so partially due to
their inherently stringy nature, as they treat the
left and right movers asymmetrically.

Irrespective of the choice of vacua, Bogolubov coefficients, hence
the global pair production rate, are given by the overlap of the
the zero-mode wave functions, or equivalently by the
two-point function of twisted fields at tree-level,
after a basis for {\it in} and {\it out} states is chosen (see
e.g. \cite{Strominger:2003fn} for a related conclusion).

We now supply some details on the vacua outlined above.

\subsubsection{Minkowski Vacuum}
In the untwisted sector one divides operators uniquely into
creation and annihilation by requiring that wave packets
built out of creation operators are analytic in the lower half plane of
the complex time coordinate. In Minkowski space, this is equivalent
to saying that the wave
functions $e^{-iEx^0}$ corresponding to creation operators have
positive energy, $E>0$. We will try and implement this
definition for the twisted sector.

For this, we return to the oscillator representation \eqref{eigo},
and analyze the analyticity properties of the wave function
in real space $f(x^+,x^-)$. For the short string
$(\epsilon,{\tilde\epsilon})=(1,1)$ case,
the kernel picks up
a converging factor $\exp(-(\a^++\a^-)0^+)$ upon continuing
$x^\pm \to x^\pm - i 0^+$: the resulting mode is thus analytic
and bounded in the lower $x^0$ half plane, hence, from the discussion
in Section 2.2, should correspond to a creation operator
in the (Minkowski) covering space vacuum. Similarly,
for $(\epsilon,{\tilde\epsilon})=(-1,-1)$ the mode is analytic
and bounded in the upper $x^0$ half plane, hence
corresponds to an annihilation operator.

On the other hand, for the long string $\eps\teps=-1$ case,
it is easy to see that the kernel $\Phi$ is analytic and bounded
neither in the upper nor in the lower half plane, hence cannot
be assigned any definite quantization rule on this basis alone.
However, recall  that the classical long string solutions are
localized within one of the Rindler wedges depending on the signs of
$\epsilon$ and ${\tilde\epsilon}$: for
$(\epsilon,{\tilde\epsilon})=(1,-1)$ they were localized in the region
$x^+>0,\ x^-<0$. This fact carries over to the quantum wave function
in a standard way - the function will decay rapidly at infinity unless
a stationary phase exists for $(\a_0^+,\talpha_0^-)$, which for
$(\epsilon,{\tilde\epsilon})=(1,-1)$ indeed happens only for $x^+>0,\
x^-<0$. Moreover, the Lorentz boost operator $j$ evolves
forward in the right Rindler region $R$, and backward in the left
Rindler region $L$, with respect to the time orientation on the
covering space. It is thus natural to split the long string
states into creation and annihilation operators according to the
sign of the product $\sgn(j) \eps$: in this way, creation operators
will indeed evolve forward in Minkowski time.

In this prescription $j=0$ states should be thought of as ``zero
modes'' of the configuration. Note that $j$ is
quantized, hence these zero modes are not part of a
continuum. Rather they are like 0+1 dimensional
quantum mechanical modes (for example String theory
compactified in all directions spatial will have
isolated vertex operators with zero momentum in all
directions, and zero energy for massless fields).
The situation here is particularly pathological
from this point of view - these modes exist for any
value of $M$, as long as ${\tilde M}^2=M^2$ and
hence can happen at for any transverse momentum and
for every oscillator number.

\subsubsection{Charged Particle Vacua}
In the oscillator representation \eqref{oscr}, the
wave function factorizes into a product of wave
functions for the left and right movers, each of
which corresponding to eigenmodes of an inverted
harmonic oscillator. As discussed in
\cite{Pioline:2003bs}, the same situation is
encountered when quantizing a particle in an
electric field in static coordinates: states
bouncing on the right of the potential barrier are
interpreted as electrons propagating forward in
time, while states bouncing on the left side of the
potential are positrons propagating backward in
time -- tunneling under the barrier being simply
Schwinger induced pair production. In addition, the
charged particle has spectator translational
zero-modes $x^\pm_0$. The closed strings modes
$\a^\pm_0$ and $\ta^\mp_0$ can thus be interpreted
as a pair of charged particles with opposite
charge, upon disregarding their translational
zero-modes.

On the basis of this analogy, one may therefore
quantize the closed string modes according to their
charged particle interpretation. Short strings with
$(\epsilon,\tilde\epsilon)=(1,1)$ correspond to the
tensor product of an electron and a positron, both
of them propagating forward in time: according to
our rule, it is thus a creation operator.
Similarly, short strings with
$(\epsilon,\tilde\epsilon)=(-1,-1)$ correspond to
the tensor product of a positron and an electron,
both of them propagating backward in time, hence
are annihilation operators. On the other hand, long
strings with $(\epsilon,\tilde\epsilon)=\pm(1,-1)$
correspond to the tensor product of two electrons
(or positrons), one of which propagating forward in
time while the other one propagates backward. This
rule therefore  does not determine the vacuum for
long strings.

However, motivated by the discussion in Section
3.3, we may break the symmetry between
the left and right movers, and choose a particular
identification of the closed string modes
$(\a_0^\pm,\ta_0^\pm)$ with the charged particle's
$(a_0^\pm, x_0^\pm)$, say relation \eqref{idop}. As
we discussed in Section \ref{opcl}, the physical
states are now eigenmodes of the Schr\"odinger
equation \eqref{schror} describing charged
particles in an electric field from the point of
view of an accelerated observer. As explained in
\cite{Gabriel:1999yz,Pioline:2003bs}, there is now
an unambiguous way to quantize the system in each
of the Rindler patches, depending on the sign of
$\eps$: in this prescription, short strings
propagating forward in time and long strings living
in the right Rindler patch are creation operators,
while short strings propagating backward in time
and long strings living in the left  Rindler patch
are annihilation operators. In order to completely
specify the vacuum, one should specify the basis of
wave functions according to their behavior at
infinity or on the light-cone, and apply the above
rule to the dominant part of the wave function. As
explained above, there is an ambiguity  in  this
prescription which of   the left or right moving
zero-modes $\a^\pm_0$ and $\ta^\pm_0$ is identified
with the charged particle's $x_0^\pm$. We see that
the two possible choices lead to two different
prescriptions for the long string modes.

\subsubsection{Quantization in the Rindler Patch \label{rindlerq}}
We now  discuss the second quantization from the
point of view of an observer sitting at a fixed
radius $r$ in one of the Rindler patches. As
emphasized in Section 3.2, for strings which
propagate in the Rindler region (be it long or
short strings), the world-sheet time $\tau$
generates a radial motion, while the $\sigma$
direction winds around the compact time coordinate
$\eta=w\sigma$. It is thus natural to first
quantize the string with respect to $\sigma$
evolution, rather than $\tau$\footnote{This
procedure should not cause any particular worry, as
it is routinely used in proving world-sheet duality
at the one-loop level.}. The energy measured by a
Rindler observer, corresponding to translations of
time $\eta$, is now given by an integral over the
non-compact $\tau$ coordinate, \be \label{rinen}
W=-\int_{-\infty}^{\infty} d\tau \left(X^+
\p_\sigma X^- - X^- \p_\sigma X^+ \right) =
\int_{-\infty}^{\infty} d\tau R^2
\partial_\sigma \eta
\ee
while the mass-shell conditions,
$M^2=2\a_0^+\a_0^-$ and its tilded counterpart, are unchanged since, at
the zero-mode level, the
world-sheet energy-momentum tensor is independent of $\sigma$ and
$\tau$, and hence leads to the same Virasoro conditions (up to an
infinite multiplicative factor) in either quantization scheme.

Notice in particular that the Rindler energy
has nothing to do with the boost momentum
$j=(M^2-\tM^2)/(2\nu)$ which was found for the
usual $\tau$ evolution, rather it is proportional
to the winding number $w$. In particular, in this
quantization scheme there appears to be no reason
to quantize the boost momentum $j$, but rather $W$
should be quantized in units of $1/\beta$.
Nevertheless, if one continues to quantize the oscillators
in the same way, the quantization of $j$ will still result
from the matching condition.

The Rindler energy \eqref{rinen}
diverges, as expected for an infinitely extended open string.
Nevertheless, it is easy to compute the Rindler energy density,
by unit of radial element,
\be
W = \int_{-\infty}^{\infty}  w(r)~\frac{dr}{d\tau}~ d\tau
\ee
where
\be
w(r) =                  \frac{4 \nu^2 r^3 \sgn(\nu)}
        {\sqrt{ (M^2+ \tM^2 - 4\nu^2 r^2) ^2 - 4 M^2 \tM^2}}
\ee This energy density diverges as
$(r-r_0)^{-1/2}$ at the turning point $r=r_0$, but
the singularity is integrable. As $r\to\infty$, for
long strings at $r\to\infty$, the energy diverges
quadratically, as a result of the infinite
tensive energy (positive for $w>0$) stored at large radius.
For short strings, one should instead integrate
over $r\in[0,r_0]$, as the string spends only a
finite interval $\tau [\tau_0,\tau_1]$ in the
Rindler patch. However it is not consistent with
energy conservation to truncate to an interval, and
indeed energy leaks from the Rindler patch to the
Milne patch. Rather, integrating over the full time
axis, one finds an infinite large negative value
(for $w>0$). In the absence of CTC, this would lead
to the conclusion that the Rindler vacuum defined
by classifying states according to the sign of $W$
is in fact unstable. It would be interesting to
understand whether the compactness of $\eta$
somehow alleviates this instability.

\section{One-loop String Vacuum Amplitude}
In this Section, we discuss  the calculation and
interpretation of the one-loop partition function.
In particular, we discuss the various divergences
and their relation to Euclidean periodic trajectories
various decay modes. The issues of particle
creation, divergent energy-momentum tensors and
back-reaction intimately depend on the choice
of vacuum.  We will be able to obtain only partial
results, with the particular choice of vacuum
implicit in the orbifold prescription of string
theory.

For orientation purposes, we start by reviewing the
case of charged open strings in an electric field,
first discussed in \cite{Bachas:bh}. We emphasize
some subtleties in the analytic continuation, both
on the world-sheet and in target space,  and argue
that the one-loop amplitude should be properly
viewed as a path integral over Euclidean world-sheet
and Lorentzian target space. We then adapt the
semi-classical methods developed in \cite{affleck}
for the ordinary, field-theoretic, Schwinger effect
to the string theory setting, recovering the
standard pair creation rate of charged open strings
from an analysis of the periodic trajectories in
imaginary time.

We then return to the subject of Misner space, and
analyze the one-loop vacuum amplitude, both in the
untwisted and twisted sectors. In the former case, while
local divergences of the energy-momentum found at the field theory
level in Section 2.4 disappear in string theory 
upon integration over space-time, 
we find new divergences (appearing for the
open string as well) arising from the sum over
Regge trajectories required in string theory. This 
genuine stringy feature should play a r\^ole
quite generally in string theory near cosmological singularities.

We then study the twisted sectors, and analyze
the imaginary periodic orbits in analogy with the
open string case. Upon regulating the singularities
of the one-loop amplitude appropriately, we find 
that the vacuum amplitude remains real, despite the
fact that the Euclidean action has instabilities
towards condensation of large winding strings. We
thus conclude that localized particle creation, and
therefore back-reaction on the geometry, does take
place in generic vacua.

\subsection{Charged Open Strings in an Electric Field}
Let us start by recalling the Schwinger computation
of pair production of charged particles in an
electric field, at the field-theory level.

\subsubsection{Analytic Continuation and Production Rate}

Using the standard Schwinger proper-time
representation \be \log \frac{a}{b} = \int_0^\infty
\frac{d\rho}{\rho} \left( e^{i\rho(b+i\eps)} -
e^{i\rho(a+i\eps)} \right) \ee the logarithm of the
vacuum-to-vacuum amplitude for a charged particle
of mass $M$, charge $\nu$  and spin $s$ can be
written as the ``MM'' integral (for Minkowski
world-sheet -- Minkowski target space) \be
\label{fschmme} {\cal F} = -\frac{1}{2(2\pi)^{D-2}}
\int_0^\infty \frac{d\rho}{\rho^{D/2}} \left[
\frac{ \nu}{\sinh(\nu \rho)} \chi_s(\rho) -
\frac{2s+1}{\rho} \right] e^{-i(M^2 - i \eps)\rho}
\ee where $\chi_s(\rho)= \sinh[(2s+1)\nu
\rho]/{\sinh(\nu\rho)}$ is the character of the
$SU(2)$ representation of spin $s$. The second term
in the bracket corresponds to the subtraction of
the one-loop vacuum amplitude in the absence of an
electric field. This integral suffers from the
usual ultraviolet divergences at $\rho\to 0$, which
give rise to a renormalization of the action of the electromagnetic
field.
Thanks to the Feynman $i\eps$ prescription, the
integral converges in the infrared $\rho\to
\infty$, however only for spin $|s|<1$. In this
case only, one may rotate the integration contour
in the upper half plane, and rewrite
\eqref{fschmme} as an ``EM'' integral (Euclidean
world-sheet, Minkowski target space), \be
\label{fscheme} {\cal F} =
-\frac{i^{D/2}}{2(2\pi)^{D-2}} \int_0^\infty
\frac{d\rho}{\rho^{D/2}} \left[ \frac{ \nu
\sin[(2s+1)\nu \rho]}{\sin^2(\nu \rho)} -
\frac{2s+1}{\rho} \right] e^{-M^2 \rho} \ee where
the contour runs slightly to the right of the poles
at $\nu \rho = k \pi$. Using the standard principal
part prescription $[x-i\eps]^{-1} =P(1/x)+ 2\pi i
\delta(x)$ , one finds that, just as the original
result \eqref{fschmme}, the dispersive (real) part
of ${\cal F}$ is given by a finite principal part
(except for the ultraviolet logarithmic divergence
at $\rho=0$), while the dissipative (imaginary)
part of ${\cal F}$, hence the pair production rate,
is a sum of residues \be \label{schpart} {\cal W}=
\Im {\cal F} =\frac{1}{2(2\pi)^{D-1}}
(2s+1)\sum_{k=1}^{\infty} (-1)^{F(k+1)}
\left(\frac{|\nu|}{k}\right)^{D/2} \exp\left( - \pi
k \frac{M^2}{|\nu|} \right) \ee Note that the
principal part prescription does not strictly
follow from Feynman's $i\eps$ prescription,
however it is necessary in order to recover the
manifestly finite result \eqref{fschmme}. Note also
that in field theory, rotating the Schwinger
parameter $\rho\to i\rho$ is essentially equivalent
to rotating the electric field   to a magnetic
field.

For spins $|s|>1$ however, the MM integral
\eqref{fschmme} is ill-defined in the infrared.  On
the other hand, the EM integral \eqref{fscheme} is
perfectly well defined, leading to a consistent
production rate \eqref{schpart}, sum of
contributions of the various helicity states, and a
finite dispersive part. Our attitude is therefore
to take the EM integral \eqref{fscheme} as the
fundamental observable, and obtain \eqref{fschmme}
by analytic continuation $\rho\to e^{i\theta}\rho$,
with $\theta$ ranging from $0$ to $\pi/2$. For spin
$|s|>1$, it is easy to see that the rotation ceases
to be valid at $\tan\theta=M^2/(2s-1)\nu$, where
the contribution at infinity does not vanish
anymore. This divergence is simply the reflection
of the Nielssen-Olesen tachyonic instability of a
massive charged particle of spin $s\geq 1$ in a
magnetic field \cite{nielsen}, which would have
been encountered if one were rotating the electric
field instead. Just as the Schwinger instability
relaxes the electric field, the Nielssen-Olesen
instability relaxes the magnetic field to zero \cite{David:2002km}.

\subsubsection{Periodic Trajectories and Schwinger Production Rate
\label{sphaleron}} We now wish to rederive the
stringy Schwinger production rate from a different
point of view, which emphasizes the semi-classical
interpretation of the process as an instanton, as
explained in \cite{affleck}. For simplicity, we
recall the argument in the case of a charged scalar
particle in an electric field. As in the Schwinger
approach, we start with the first-quantized
representation of the free-energy
(i.e. the logarithm of the
vacuum-to-vacuum transition amplitude), \be
\label{free} \mathcal{F}= -\int_0^{\infty} \frac{d\rho}{\rho}
\int [DX(\tau)]~ e^{-S[X]}
\ee
where $S$ is the action
for a particle of charge $\nu$ in an electric
field in imaginary proper time,
\be \label{action}
S[X] = 2\pi M^2 \rho+ \int_0^{2\pi\rho}
\left[ \partial_\tau X^+
\partial_\tau X^-
-\frac{i\nu}{2}\left(X^+\partial_\tau X^- -
X^-\partial_\tau X^+\right)  \right] d\tau
\ee
The path integral is restricted to configurations $X(\tau)$
periodic in imaginary time $\tau$ with period $2\pi\rho$ \footnote{We
deviate from the notation in the previous subsection, in order
to match the notation in the closed string computation.}.

 Schwinger's computation consists in evaluating the
path integral over $X(\tau)$ first, obtaining
$\mathcal{F}(\rho)$, and then integrate over the
proper time $\rho$  to obtain $\mathcal{F}= -
\int_0^\infty \frac{d\rho}{\rho} \mathcal{F}(\rho)
$. By the optical theorem, the total decay rate of
the vacuum is given by the imaginary part of the
free energy. Since the world-line action is real,
an imaginary part can only arise from singularities
in $\mathcal{F}(\rho)$. However, not all such
singularities of $\mathcal{F}(\rho)$ are related to
decay processes, for example we will see that
singularities in $\mathcal{F}(\rho)$ may also arise
from the rapid growth of the density of states, as
is the case for the stringy Hagedorn behavior.
Therefore, to avoid confusion, we will relate
directly periodic classical solutions to decay
rates, bypassing whenever possible a discussion of
$\mathcal{F}(\rho)$ and its singularities.

Instead, as proposed in  \cite{affleck}, one may first carry out
the integration over $\rho$, which is of Bessel type after rescaling
$\tau\to\rho\tau$. In the saddle point (large $M$) approximation,
the integral is peaked around
\be
\rho=\frac{1}{M} \sqrt{\int_0^{2\pi} d\tau \left(\partial_\tau
X^+\partial_\tau X^-\right) }
\ee
and yields the saddle point action,
\be
\label{free2}
S= 2 M
\sqrt{\int_0^{2\pi} d\tau \left(\partial_\tau
X^+\partial_\tau X^-\right) }-
\frac{i\nu}{2}\int_0^{2\pi} d\tau\left(X^+\partial_\tau
X^- - X^-\partial_\tau X^+\right)
\ee
The path integral is now over periodic configurations
in imaginary time with period 1, and can be evaluated
semi-classically by expanding around the classical solutions.
Despite the apparent non-linearities in \eqref{free2}, classical solutions
are still the usual charged particle trajectories, up to a rescaling of
imaginary time $\tau\to\rho\tau$,
\be \label{cs1}
X^\pm_0(\tau) = x_0^\pm \pm
\frac{{a_0^\pm}}{\nu} e^{\pm i k \tau}
\ee
where the integer $k$ indicates the number of revolutions during
the period $2\pi$. The radius of the circle $R^2:=2 a^+ a^- $
is fixed by the
equations of motion resulting from \eqref{free2} to the on-shell
value $R=M$, while the classical action and period
are given by
\be
S_k = \pi k M^2/ \nu\ ,\quad \rho_k = k/\nu
\ee
Notice that $\rho_k$ is precisely the value of the
Schwinger parameter where the one-loop amplitude
\eqref{schpart} becomes singular.

In order to evaluate the path integral, one must
now compute the determinant of fluctuations around
\eqref{cs1}. The complete analysis was carried out
in \cite{affleck}, where it was found that
fluctuations of radius $R$ of the trajectory around
the saddle value $R=M$ corresponds to a negative
eigenvalue for the action, while other deformations
of the trajectory have a positive eigenvalue. In
addition, the zero-mode $x_0^\pm$ of course has zero
action and leads to a factor of volume, as
appropriate for the total vacuum-to-vacuum
transition amplitude. For our purposes, it will be
sufficient to concentrate on the radial mode $R$.
Evaluating the action \eqref{free2} (before any
rescaling of time) on the off-shell configuration
\eqref{cs1}, we obtain \be S[X_0] = \frac{\pi
k^2}{\nu^2 \rho} \left( 1 - \frac{\nu \rho}{k}
\right) R^2 + \pi M^2 \rho \ee Clearly, when
$\rho=k/\nu$ the Gaussian integral over $R^2$
diverges, which is the origin of the divergence in
\eqref{schpart}. However, one may integrate over
$\rho$ first, and replace it by its saddle point
value  $\rho=|k| R /(\nu M)$, obtaining \be
\label{afef} S = \frac{k \pi}{\nu} R (2M-R) \ee
This action indeed has a maximum at $R=M$, leading
to an imaginary part for the transition amplitude.
The sum over the Euclidean solutions \eqref{cs1}
for all $k$, including the one loop determinant,
generates the classic Schwinger result for the
vacuum decay rate \eqref{schpart} \cite{affleck}.

This semi-classical derivation of the spontaneous
production rate is complementary to the semi-classical
derivation of induced pair production outlined
in Section 3.3. Both processes involve a propagation
in imaginary proper time: in the case of induced pair production,
the Lorentzian trajectory is continued to Euclidean
time at the turning point, evolved through the
potential barrier, and continued back to Lorentzian
on the other side. In the case of spontaneous pair production, an
Euclidean periodic trajectory is cut at
mid-period and used as an initial electron-positron state
which undergoes subsequent Lorentzian evolution (much as in the
Hartle-Hawking prescription for quantum gravity).
Since the zero modes $x^\pm_0$
are arbitrary, the pair production is spatially
homogeneous.
As we shall see, these two semi-classical pictures  carry over to the
closed string case, with appropriate adjustments.

\subsubsection{Charged Open Strings in an Electric Field}

We now briefly proceed to the case of open strings
stretched between two D-branes carrying different
uniform electric fields $e_0,e_1$ \cite{Bachas:bh,elecfield}. Such open
strings carry an overall charge under the electric
field $e_0-e_1$, hence behave much like charged
particles, except for additional polarization
effects. In field theory, a standard assumption is that
the Schwinger proper time can be rotated from Minkowski to Euclidean
without difficulties. We will see that the excited string modes
invalidate this lore in string theory.

Starting with the semi-classical approach outlined
above, we look for periodic classical trajectories
in imaginary world-sheet time. A complete set of
classical  solutions with those boundary conditions
is given by
\be \label{xn} X^\pm_n (\sigma,
\tau) = e^{(n\pm i\nu) \tau} \cos \left[(n\pm
i\nu)\sigma \right] \ee
where $\pi \nu=\arctan(\pi e_0)
+\arctan(\pi e_1)$ \cite{Bachas:bh}.
Unless $n=0$, none of these
solutions is periodic in Euclidean time. The mode
$n=0$ corresponds to the same classical solution
\eqref{cs1} for the open string center of motion,
therefore the partition function has only the
previous set of saddle points. The contribution of
the stringy modes to the one loop determinant
around each of them can be extracted from the
computation by Bachas and Porrati of the one-loop
vacuum amplitude, e.g. for the bosonic open string
\be \label{abos} A_{bos}=  \frac{i \pi V_{26} (
e_0+ e_1) }{2} \int_0^\infty \frac{d\rho}{(4\pi^2
\rho)^{13}} \frac{e^{-\pi\nu^2
\rho/2}}{\eta^{21}(i\rho/2)~\theta_1(\rho\nu/2;
i\rho/2)} \ee where $\theta_1$ is the Jacobi
theta function, \be \label{jac} \theta_1(v;\rho)= 2
q^{1/8} \sin \pi v \prod_{n=1}^\infty (1-e^{2\pi i
v} q^n) ( 1-q^n) (1-e^{-2\pi i v} q^n) \ ,\quad
q=e^{2\pi i\rho} \ee As in the field-theoretical EM
integral above, the poles on the real $\tau$ axis
at $\rho=2m/\nu$ contribute to the imaginary part,
yielding a sum of the Schwinger pair creation rates
\eqref{schpart} over excited states of the open
string spectrum (with a mass shift $M^2\to
M^2+\nu^2$, in the bosonic string case only). This
is the result found in  \cite{Bachas:bh}.

On the other hand,  the partition function has
additional poles in the complex $\rho$ plane, at
$\rho\nu/2=m+n~i\rho/2$. If we were to adopt the
prescription of starting  from a Minkowskian
world-sheet and then rotating $\rho\to i\rho$, these
poles would give an extra contribution to the
creation rate, equal to \be \sum_{m,n}
e^{-i\pi\nu^2 \rho/2} \eta^{-24} \left[
\frac{m}{n+i\nu} \right] \ee This prescription
would however be in disagreement with the
semi-classical reasoning outlined above. We
conclude that the Schwinger production rate should
not include the contributions of the poles away
from the imaginary proper time axis.
Thus, computing the imaginary part using the imaginary and the real
Schwinger proper time contours gives different results. The two are
not equivalent and the additional string poles, associated with
excited strings modes, invalidate the standard field theory procedure
of analytically continuing from one to the other.

As in the charged particle case, it is possible
to understand the origin of the poles at the level of the path integral:
For $\rho\nu/2=m+n~i\rho/2$, the action corresponding to the excited mode
\eqref{xn} vanishes, and the integral over $a_n^\pm$ diverges.
It is also possible to regard these poles
as a consequence of the existence of Regge
trajectories in the string spectrum. Indeed,
starting with the Schwinger result \eqref{schpart}
for fields with arbitrary spin $S$, we see that the
sum over an infinite set of states with $M^2=M_0^2+
n S$, $S=0\dots\infty$ gives rise to poles in the
complex $\rho$ plane at $\rho \nu/2= k + n
i\rho/2$. These are precisely the extra poles of
the partition function, arising from the Jacobi
theta function in the denominator of \eqref{abos}.
A similar effect will be shown to occur shortly in
the closed string case. However, for the
calculation of the pair production rate, these
divergences are irrelevant. Indeed the production
rate for each particle species depends on the spin
through an overall multiplicity factor of $2s+1$,
hence remains finite.

To summarize, the lesson is that the decay rate can
be read off from the one-loop vacuum amplitude with
an Euclidean world-sheet, and a target space
signature fixed by the existence of periodic
orbits. Poles away from the imaginary world-sheet
time axis do not contribute to the pair creation
rate, and can be viewed as a consequence of the existence of
states with arbitrarily high spin in the spectrum.

 \subsection{Closed String Theory Vacuum Amplitude}
After this detour via open strings in an electric fields, we now
return to closed strings in Misner space, and analyze the one-loop
vacuum amplitude, computed by path integral
and Hamiltonian methods in \cite{Nekrasov:2002kf,Cornalba:2002fi}.
The one-loop vacuum amplitude can 
be written as an ``EM'' (Euclidean world-sheet,
Minkowskian target space) integral on the fundamental domain
of the moduli space of the torus,
\be
\label{milne1l} A_{bos}=\int_{{\cal F}}
\sum_{l,w=-\infty}^{\infty} \frac{d\rho_1
d\rho_2}{(2\pi^2 \rho_2)^{13}} \frac{e^{-2\pi
\beta^2 w^2 \rho_2 -
\frac{R^2}{4\pi\rho_2}|l+w\rho|^2}} {\left|
\eta^{21}(\rho) ~\theta_1\left[i \beta(l+w\rho); \rho\right]
\right| ^2 } \ee
where $\rho=\rho_1+i\rho_2$, and 
$\theta_1$ and $\eta$ are the Jacobi
and Dedekind functions, \bea \label{jac} \theta_1(v;\rho)&=& 2
q^{1/8} \sin \pi v \prod_{n=1}^\infty (1-e^{2\pi i
v} q^n) ( 1-q^n) (1-e^{-2\pi i v} q^n) \\
\eta(\rho)&=&q^{1/24}  \prod_{n=1}^\infty (1-q^n)\ ,\qquad q=e^{2\pi i\rho}
\eea
Expanding the inverse of the
Jacobi function in powers of $q$, one easily recovers
the spectrum of the world-sheet conformal field theory. In
particular, as explained in \cite{Pioline:2003bs}, the zero-mode contribution
$|\sinh(\beta(l+w\rho)]|^2$ can be interpreted either as the
contribution of the (left-moving and right-moving)
discrete spectrum of the harmonic oscillator
in Euclidean time, or as the contribution of the continuous
real spectrum of the inverted harmonic oscillator in Lorentzian
time.

In this Section, we analyze the target space
interpretation of the one-loop amplitude
\eqref{milne1l}, with particular attention to the
singularities of the integrand in the moduli space
of the torus, at \be \label{div}\rho = \frac{k- i
\beta l}{n + i \beta w} \ee where the integer $n$
labels which factor in the product formula of the
Jacobi theta function vanishes, and the integer $k$
parametrizes the poles. In contrast to the open
string case, these poles occur simultaneously on
the holomorphic and anti-holomorphic side, and
therefore are not expected to yield an imaginary part, rather,
they give rise to logarithmic divergences. Such
singularities are very reminiscent of the long
string poles appearing in Euclidean AdS$_3$
\cite{Maldacena:2000kv}, and, as we shall see, can
also be related to infrared problems.

\subsubsection{Untwisted sector}

We start with the untwisted sector $w=0$. Expanding in powers
of $q$ the inverse of the Dedekind function and the $n\neq 0$ terms of the
Jacobi functions, it is apparent that the string theory
vacuum amplitude can be viewed as the field theory
result \eqref{ft} summed over the spectrum of
(single particle) excited states, satisfying the
matching condition enforced by the integration over
$\rho_1$. As usual, field-theoretical UV
divergences at $\rho\to 0$ are cut-off by
restricting the integral to the fundamental domain
$F$ of the upper half plane. By the standard unfolding
trick, one may unfold the integration domain to the 
strip $[0,1]+i\Real^+$ upon restricting the sum to
$w=0$.

In contrast to the field theory result, where the
integrated free energy is finite for each particle
separately, the free energy here has  poles in the
domain of integration, at \be \rho= \frac{k-i\beta
l}{n}  \ee Those poles arise only after summing
over infinitely many string theory states. Indeed,
each pole originates from the $(1-e^{\pm 2\pi i v}
q^n)$ factor in \eqref{jac}, hence re-sums the
contributions of a complete Regge trajectory of
fields with mass $M^2= M_0^2+\alpha' p$ and spin $s=p$
($p\in\Nint$). In other words, the usual
exponential suppression of the partition function
by the increasing masses along the Regge trajectory
is overcome here by the spin dependence of that
partition function. Regge trajectories are a
universal feature of perturbative string theory,
and we believe that this distinctive stringy effect
should manifest itself generically in the presence
of space-like singularities. From this point of
view, the poles \eqref{div} with $w=0$ are related
to ultraviolet divergences due to higher spin
effects.

Finally, notice that the dependence on the
zero-modes $(x^+,x^-)$ could be reinstated in the
string one-loop amplitude by writing the zero-mode
part of the Jacobi function $\sinh^2 \pi\beta l$ as
a Gaussian integral similar to \eqref{ftg}. This
suggests that the one-loop energy momentum, probed
by on-shell vertex operators, will again diverge on
the chronology horizon \cite{bdpr}. While this divergence may
be somewhat alleviated in the superstring case, we
still expect divergences as supersymmetry is
explicitly broken by the cosmological background.

\subsubsection{Twisted sectors}
Let us now turn to the twisted sectors, with $w\neq
0$. The target space interpretation of the one-loop
amplitude \eqref{milne1l} can be understood by
adapting  the computation of the charged particle
path integral, outlined in Section 4.1.2, to the
closed string case. Following the general lesson
from section 4.1, we take the target space to be
Lorentzian ($X^\pm$ real), but rotate the
worldsheet time to the imaginary axis. The closed
string path integral is given by \be \mathcal{F} =
\int_F \frac{d\rho_1 d\rho_2}{\rho_2} \sum
_{l,w=0}^\infty \int [DX] e^{-S_{l,w}} \ee with
action (for $\a' = 1$)  \be \label{ac}
S=\frac{1}{2} 
\int_0^{2\pi\rho_2} d\tau
\int_0^{2\pi} d\sigma \left( \pa_\tau X^+ \pa_\tau
X^- + \pa_\sigma X^+ \pa_\sigma X^- \right) \ee The
path integral is restricted to periodic
configurations on the torus of modulus
$\rho=\rho_1+i \rho_2$, \bea X^\pm (\sigma +2\pi,
\tau) = e^{\pm 2\pi\beta w} X^\pm
(\sigma , \tau) \nonumber \\
X^\pm (\sigma + 2\pi\rho_1, \tau+ 2 \pi \rho_2) =
e^{\mp 2\pi\beta
 l} X^\pm (\sigma , \tau)
\eea where $(-l,w)$ denote the orbifold twists in
the $\sigma$ and $\tau$ directions. 
Interpreting the path integral as the sum of
single particle free energies, the integration over
$\rho_1$ enforces level matching, while $\rho_2$ is
interpreted as the Schwinger parameter appearing in
the previous discussion, with a UV cut-off
enforced by the restriction to the fundamental domain $F$

For simplicity, we now restrict the path integral
to quasi-zero-mode configurations (which become
solutions only when $\rho$ coincides with ones of the poles), 
\be \label{qzm}
X^\pm = \pm \frac{1}{2\nu} \alpha^\pm e^{\mp (\nu
\sigma - i A \tau)} \mp \frac1{2\nu} \talpha^\pm
e^{\mp (\nu\sigma + i \tilde A\tau)} \ee where
$\nu=-\beta w$ and \be A= \frac{k}{\rho_2} - i
\beta \frac{l + \rho_1 w}{\rho_2} \ ,\quad \tilde
A= \frac{\tilde k}{\rho_2} + i \beta \frac{l +
\rho_1 w}{\rho_2} \ee and $k,\tk$ are a pair of
integers labelling the periodic trajectory, for
fixed twists $(l,w)$. In order to satisfy the
reality condition on $X^\pm$, one should restrict
to configurations with
\be
\label{rc1}
k=\tilde k\ ,\qquad \alpha^\pm = - (\talpha^\pm)^*
\ee
Nevertheless, for the sake of generality we shall not
impose these conditions at this stage. Notice
then that for $k=-\tk$, the world-sheet described by \eqref{qzm}
becomes degenerate, and half of the modes, say $\ta^\pm$, 
should be set to zero to avoid redundancy. This case will
be excluded in the following discussion.

We now evaluate the Euclidean world-sheet action \eqref{ac}
for such a configuration,
\bea
S&=&  \frac{\pi }{2\nu^2 \rho_2}
\left( \left[k - i (\beta l + \nu \rho_1)\right]^2 - \nu^2 \rho_2^2 
\right) ~\a^+ \a^-
+\frac{\pi }{2\nu^2 \rho_2}
\left( \left[\tk + i (\beta l + \nu \rho_1)\right]^2 -\nu^2 \rho_2^2 - 
\right) ~\ta^+ \ta^- \nn \\
&&- 2 \pi i j \nu \rho_1 + 2 \pi  \mu^2 \rho_2
\eea
where the last line, equal to $-i\pi \rho M^2 +i \pi \tilde \rho \tM^2$,
summarizes the contributions of excited modes. It is then convenient 
to use hyperbolic angular coordinates,
\be
\a^\pm = \pm R e^{\pm \eta}/\sqrt{2} \ ,\qquad
\ta^\pm = \pm \tR e^{\pm \tilde \eta}/\sqrt{2}
\ee
where the reality condition imposes 
\be\tilde\eta=\eta^*\ ,\quad
R^2=\tilde R^2 \label{rc2}\ee
The integration over the overall angle
$\eta+\tilde\eta$ gives a finite constant $2\pi\beta$ due to the 
orbifold identification, while 
the integration over $\eta-\tilde\eta$ gives an infinite factor, independent
of the moduli.

In addition, there are divergences coming from the
integration over $R$ and $\tilde R$ whenever either
of the two following conditions are satisfied,
\be\label{pp}
\nu \rho_2 = \pm  \left[k - i (\beta l + \nu \rho_1)\right] \qquad
\mbox{or}\qquad
\nu \rho_2 = \pm  \left[\tilde k + i (\beta l + \nu \rho_1)\right]
\ee
(with uncorrelated signs) The two conditions
above are then satisfied simultaneously at
\be
\label{r12}
\rho_1=-\frac{\beta l}{\nu} - i \frac{\nu}{2} (k - \tilde k)\ ,\qquad
\rho_2 = \frac{|k+\tk|}{2\nu}
\ee
which, for $k=\tilde k$, are precisely the double poles \eqref{div}.
We can therefore interpret these poles as coming from infrared divergences
due to existence of modes with arbitrary size $(R,\tR)$. For $k\neq
\tilde k$, the double poles are now in the
complex $\rho_1$ plane, and may contribute for specific
choices of integration contours, or second-quantized vacua.
In either case, these
divergences may be regulated by enforcing a cut off
$|\rho-\rho_0|>\eps$ on the moduli space, or an infrared cut-off on $R$.
It would be interesting to understand the deformation of Misner
space corresponding to this cut off, analogous to the Liouville
wall in $AdS_3$ \cite{Maldacena:2000kv}.

Rather than integrating over $R,\tR$ first, which is ill-defined at
$(\rho_1,\rho_2)$ satisfying \eqref{pp}, we may choose to integrate
over the modulus $\rho,\tilde\rho$ first. The integral with respect to
$\rho_1$ is Gaussian, dominated by a saddle point at
\be
\rho_1 = - \frac{\beta l}{\nu}+ i \frac{\tilde k \tilde R^2
- k R^2}{\nu (R^2+ \tilde R^2)}
- 2 i \frac{j \nu \rho_2}{R^2+\tilde R^2}
\ee
It is important to note that this saddle point is a local maximum
of the Euclidean action, unstable under perturbations of $\rho_1$.
The resulting Bessel-type action
\be
S = \pi \rho_2 \left[ 2\mu^2 - \frac12 (R^2 + \tilde R^2)
-\frac{j^2 \nu^2}{R^2 + \tilde R^2} \right]
+ \frac{\pi}{2\nu^2\rho_2}
\frac{(k+\tilde k)^2R^2 \tilde R^2}{R^2+\tilde
  R^2}
+ 2\pi  j \frac{k R^2 - \tilde k \tilde R^2}{R^2 +
\tilde R^2} -2\pi i \beta j l \ee 
has again, for $k+\tk\neq 0$, a (now stable) saddle point in $\rho_2$,
at \be \rho_2 = \frac{R \tilde R
|k + \tilde k|} {\nu
\sqrt{(R^2+\tR^2)(4\mu^2-R^2-\tR^2)-4 j^2\nu^2}}
\ee 
Integrating over $\rho_2$ in the saddle point
approximation, we finally obtain the action as a
function of the radii $R,\tR$: \be \label{s2}
S=\frac{|k+\tilde k| R\tR
  \sqrt{(R^2+\tR^2)(R^2+\tR^2-4\mu^2)+4j^2\nu^2}}
{\nu(R^2+\tR^2)}
\pm 2\pi  j \frac{\tilde k \tR^2-k R^2}{\nu(R^2+\tR^2)} \pm 2\pi i
\beta j l
\ee
where the sign is that of $k+\tk$.
In analogy with the open string action \eqref{afef}, we may now
analyze the behavior of this action as a function of $R$ and $\tR$.
Not surprisingly, we find that it admits an extremum at
the on-shell values
\be
\label{sos}
R^2=\mu^2-j\nu, \qquad \tR^2=\mu^2+j\nu
\qquad \mbox{with action} \quad
S_{k,\tk}= \frac{\pi M \tM }{\nu} |k+\tilde k|
\ee
Notice that these values are consistent with the reality condition
\eqref{rc2}, since the boost momentum $j$ is imaginary in Euclidean
proper time. Evaluating $(\rho_1,\rho_2)$ for the values \eqref{sos}, 
we reproduce \eqref{r12}, which implies that the integral is 
indeed dominated by the region around the double pole. 
Under variations of $R$ and $\tR$ around the on-shell values
\eqref{sos}, we find that the Euclidean action varies as
\be
\Delta S = \frac{\pi (k+\tk)}{\nu} \left[
(\Delta R)^2 + (\Delta \tR)^2 \right]
= \frac{2\pi (k+\tk)}{\nu} \left[
(\Delta R_1)^2 - (\Delta R_2)^2 \right]
\ee
where $R=R_1+iR_2, \tR=R_1-i R_2$, due to the reality condition
\eqref{rc2}. Together with the negative eigenvalue along $\rho_1$
and the positive one along $\rho_2$, we find that
fluctuations  in $(\rho_1,\rho_2,R,\tR)$ directions around
the saddle point have  signature
$(+,+,-,-)$, hence a positive fluctuation determinant\footnote{As in
  the open string case, it is plausible to assume that 
higher excited modes do not provide further destabilizing modes,
at least in the vicinity of poles such as \eqref{div} with $n=0$. },
equal to $M^2 \tM^2$ up to a positive numerical constant.
This implies that the one-loop amplitude in the twisted sectors
does not have any imaginary part, in accordance with the naive
expectation based on the double pole singularities. It is also
in agreement with the answer in the untwisted sectors, where
the globally defined $in$ and $out$ vacua where shown to be
identical, despite the occurence of pair production at
intermediate times. 

Nevertheless, the instability of the Euclidean action under
fluctuations of $\rho_1$ and $\rho_2$ 
indicates that spontaneous pair production
takes place, by condensation of the two unstable modes.
Thus, we find that winding string production 
takes place in Misner space, at least
for vacua such that the integration contours picks up contributions
from these states. This is consistent with our discussion of the
first quantized wave functions, where tunneling  in the Rindler
regions implies induced pair production of short
and long strings. The periodic trajectories \eqref{qzm}
describe the propagation across the potential barrier in imaginary
proper time, and correspond to Euclidean world-sheet interpolating
between the string Lorentzian world-sheets.

\section{The Effects of Winding Strings}
Our analysis so far has concentrated on the
specific case of the Lorentzian orbifold, for which
we have shown that winding strings were produced in
generic vacua by a tunneling process, analogous to
Schwinger pair production of charged particles in
an electric field. The all-important question is to
determine whether the back-reaction resulting from
the production of these states is sufficient, in
specific cases, to resolve the cosmological
singularity.

As a first step on this direction we investigate
the effects of winding string in a class of
cosmological backgrounds with a single compact
direction, which includes the Misner space. In
Section 5.1, we study this problem at a classical
level, assuming that a constant number of strings
winding along the compact direction have condensed.
We find that indeed, for low transverse
dimensionality and certain initial conditions, the
compact direction may experience a bounce, rather
analogous to the usual BKL behavior in
inhomogeneous spaces \cite{BKL}.

 In Section 5.2, we
assume that the back-reaction is described by a
mean field deformed geometry, and compute the
production rate by generalizing the arguments of
Section 3. We find that, for a smooth geometry, the
production rate is suppressed at high winding
number, hence curing the pathology of the singular
Misner space. We leave it as a challenging open
problem to determine the mean field geometry
self-consistently.

\subsection{Back-reaction of Winding Strings}
One of the main  reasons to  believe that the
production of winding strings may resolve the
singularity is that, once  produced, winding
strings contribute an energy proportional  to the
radius of the  compact dimension, hence acting like
a two-dimensional positive cosmological constant.
It is thus plausible that the resulting transient
inflation may be sufficient to drive a bounce in
the geometry and avoid the
singularity\footnote{Positive spatial curvature is
usually necessary along with positive $\Lambda$ to
achieve a bounce, however two-dimensional de-Sitter
space provides a counterexample. See e.g. \cite{Molina-Paris:1998tx}
for a lucid discussion of energy conditions and cosmological bounces.}.
At first sight,
this may seem to go opposite to the mechanism
proposed in
\cite{Brandenberger:1988aj,Tseytlin:1991xk} where
the condensation of winding strings was invoked to
stop the expansion of the universe beyond three
spatial dimensions -- while momentum states
prevented the collapse to zero size. However, the
analysis in
\cite{Brandenberger:1988aj,Tseytlin:1991xk} relies
on a isotropic distribution of winding strings,
whereas we will rely on anisotropies. Furthermore,
these authors assumed the presence of  both winding
and momentum states (as dictated by the condition
of thermal equilibrium), whereas in our case
momentum states become very heavy towards the
singularity and are thus unlikely to be produced.
Somewhat surprisingly, we will find that our bounce
 takes place only when the number of spatial
dimensions is less or equal to 3.

\subsubsection{Homogeneous Kasner solutions with matter}
To investigate this mechanism, let us consider a
homogeneous Kasner-type solution \be ds^2 = -dT^2 +
\sum_{i=1}^{D} a_i^2(T) dx_i^2\ ,\qquad \ee of
Einstein's  equations with a general conserved
energy momentum tensor source, \be T_{\mu\nu}  =
\mbox{diag}( \rho, a_i^2 ~p_i ) \ee where $\rho$
and $p_i$ are the total energy density and pressure
along the direction $i$, respectively. Defining the
Hubble parameters as $H_i=\dot{a_i}/a_i$ and
setting the Newton constant to $1/(4\pi)$, the
Friedmann equation reads \be \label{fr} \rho =
\sum_{i<j} H_j H_j \ee The equations of  motion for
the radii can be written as \be \label{dife}
\dot{H_i} = -H_i \left(\sum_{j=1}^D H_j \right) +
p_i + \frac{1}{D-1} \left( \rho - \sum_{j=1}^D p_j
\right) \ee The consistency of (\ref{fr}) and
(\ref{dife}) requires the conservation law for the
energy-momentum tensor, \be \label{cons} \dot \rho
+ \sum_i H_i (p_i+\rho) = 0 \ee The evolution of
the volume element of the spatial directions is
easily obtained by summing \eqref{dife} over all
directions, \be \label{difet} \dot{H} = -H^2  +
\frac{1}{D-1} \left( D \rho - \sum_{i=1}^D p_i
\right) \ee where $H=\sum_{i=1}^D H_i$.

In the absence of matter, i.e. when $\rho=p_i=0$,
the general solution is the well-known Kasner class
of geometries, where each radius evolves
monotonously as a power law, \be a_i(T)  \sim
T^{\alpha_i}\ ,\qquad \sum_{j=1}^D \a_j =
\sum_{j=1}^D (\a_j)^2 = 1 \ee Except for the Misner
case $(1,0,0,\dots)$, at  least one of the exponents
$\a_i$ must be negative, and another two must be
positive. The Hubble parameters $H_i=\a_i/t$ blow
up at $t=0^-$, where all radii shrink to zero size
simultaneously.

In the presence of spatial inhomogeneities, these
Kasner epoches are separated by bounces, where one
shrinking direction becomes expanding while another
expanding direction turns into a shrinking one
\cite{BKL}, while the overall volume keeps
decreasing towards $t=0$. Here we restrict to
homogenous cosmologies only, so are not interested
in these kind of bounces.

\subsubsection{General Conditions for a Bounce}

Returning to a general conserved energy-momentum
tensor, we can now read from \eqref{dife} a
necessary condition for a bounce to occur in the
$i$-th direction: the time derivative $\dot{ H_i}$
should be strictly positive at $H_i=0$, that is
when the extremal size is attained. Hence \be
\label{bouc} \rho \geq \sum_{j=1}^D p_j - (D-1) p_i
\ee or, more covariantly, $\eta^{\mu\nu} T_{\mu\nu}
< (D-1) g^{ii} T_{ii}$. This condition is generally
easy to satisfy, for example by using pressure-less
dust of positive energy density. In string theory,
a gas of massless gravitons (momentum modes) in 10
dimensions will satisfy this condition.

For completeness we discuss the conditions for a
bounce of the overall volume. This  is much more
difficult to achieve: demanding that $\dot H>0$
when $H=0$, and noting that (\ref{fr}) can be
rewritten as $\rho=(H^2 - \sum_{i=1}^D H_i^2)/2$,
we get from (\ref{fr}) and (\ref{difet}) the
conditions, \be \label{bouco} \rho \leq 0~~~~
\mbox{and}~~~~ (D-2)\rho + \sum_{j=1}^D p_j \leq 0
\ee The first of these conditions (on $H$) is
fairly restrictive, but can be possibly relaxed
e.g. by including coupling to the dilaton in weakly
coupled string theory. The second condition (on
$\dot{H}$) can be easily satisfied for $D\leq2$.
Even if this condition is not sufficient to
guarantee the existence of a bounce, it is
important to note that it implies a positive
contribution to the r.h.s. of (\ref{difet}), and
thus will delay the appearance of the singularity
$H\to -\infty$.

The simplest situation  to    consider is an
isotropic distribution of matter, and all $H_i$
equal. In this case  we can derive the equality\be
\dot{H_i}= -\frac{1}{D-1} (p_i +\rho)\ee We see
that the condition on $\dot{H_i}$ requires
violation of the null energy condition in the
isotropic case, whereas the situation improves
using anisotropies.

In the following, returning to  our discussion of
Misner space and its deformations, we consider a
case where one direction $i$ is compact, and the
only non-vanishing components of the energy
momentum tensor are $T_{00}=\rho$ and $T_{ii}=a_i^2
p_i$. The bounce condition (\ref{bouc}) on the
$i$'th direction becomes $\rho\geq (2-D) p_i$. A
momentum mode along direction $i$, with $\rho=p_i$,
is thus no longer able to induce a bounce. A
winding mode along direction $i$ on the other hand
may be modeled by an energy momentum tensor \be
\rho = N \frac{{\cal T}_2}{V}\ ,\quad p_i=-\rho\
,\quad p_j=0\ ,\quad j\neq i \ee where
$V=\prod_{j\neq i} a_j$ is the volume of the
non-compact directions. From (\ref{bouco}) we see
that the equation of state $\rho=-p_i$ can lead to
a bounce in the direction $i$ if the total number
of spatial dimensions is $D\leq 3$ (although the
total volume will continue decreasing, as discussed
above).

This condition may in fact  be strengthened by
noting that, under the assumption that only
$T_{00}$ and $T_{ii}$ are non-vanishing and that
the geometry is isotropic in the remaining $(D-1)$
spatial directions $j\neq i$, one may eliminate the
pressure $p_i$ and energy density $\rho$  from the
equation of motion, and obtain \be {\dot
H}_i+(D-2){\dot H_j}=-H_i^2-(D-2)H_i
H_j-{(D-1)(D-2)\over 2} H_j^2 \ee For $D>2$, the
r.h.s. is a negative definite quadratic form, and
thus the quantity\footnote{Note this is not the
overall volume, which is instead $a_i a_j^{D-1}$.}
$a_i a_j^{D-2}$, once decreasing, keeps decreasing
until a singularity is reached. For $D\leq 2$,
however, it is possible to find solutions where a
bounce does take place.

\subsubsection{Coupling to the Dilaton}

Of course, winding strings couple to  the dilaton,
and it is not legitimate to neglect the running of
the dilaton. Its effect however can be easily
introduced by viewing it as the radius of an extra
spatial compact direction $k$, and the winding
strings  as membranes wrapped on the $(i,k)$ torus.
In units of the eleven-dimensional Planck length
$l_M$, the energy-momentum tensor describing this
situation is \be \label{tem} \rho=N\frac{{\cal
T}_3}{V}\ ,\quad p_i = p_k = -\rho\ , \quad
p_{j}=0\ , \quad j\neq (i,k) \ee where now
$V=\prod_{j\neq (i,k)} a_j$. The condition
(\ref{bouc}) now indicates that the bounce occurs
only when $D \leq 4$, when $D$ is still the total
number of spatial dimensions (including the 11-th
one). The critical dimension for a bounce induced
by the condensation of winding strings thus appears
to be $3+1$.

Again, one may strengthen this condition  by
assuming that only $T_{00}, T_{ii}$ and $T_{kk}$
are non-vanishing and that the geometry is
isotropic in the remaining $(D-2)$ spatial
directions $j\neq (i,k)$: upon eliminating the
pressures $p_i, p_k$ and the energy density $\rho$
from the equations of motion, one obtains \be \dot
H_i + (D-3) \dot H_j + \dot H_k = - H_i^2 - H_k^2 -
H_i H_k - (D-3) H_j(H_i+H_k) -\frac12(D-2)(D-3)
H_j^2 \ee Again, for $D>3$ the r.h.s. is negative
definite, and the combination $a_i a_k a_j^{D-3}$
  will keep on
decreasing monotonously if it starts out
decreasing.

\begin{figure}\label{bfig}
\begin{center}\includegraphics[height=10cm]{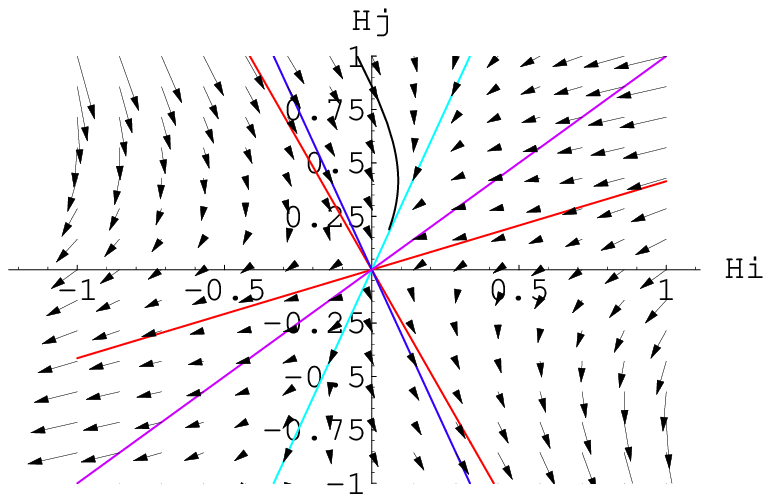}
\end{center}
\caption{Cosmological flow in the presence of a
condensate of winding strings along direction $i$,
for  $D=3, \mu=1/2$. The axes denote the Hubble
parameters of the compact direction $i$ and
non-compact directions $j$ in 11D Einstein frame.
The dark blue line is the vanishing locus for the
Hubble parameter of the direction $i$ in 10D string
frame, i.e. $H_i + (H_k/2)=0$. The purple line is
the vanishing locus for the  Hubble parameter for
the 11th direction $H_k$,
obtained from Eq. (5.14). 
The trajectory shown in black corresponds to a
bounce of the compact direction $i$ both in 11D
Einstein and 10D string frame. It asymptotes to the
clear blue line, corresponding to the fixed point
$O$. }
\end{figure}

In order to make sure that a bounce of the single
compact direction does indeed take place, let us
study in more detail the solution corresponding to
the energy momentum tensor (\ref{tem}). Since the
differential system (\ref{dife}) is homogeneous of
degree  2, it is useful to define $x=H_j/H_i$ and
$y=H_k/H_i$. For these variables, the system
reduces to \be \label{xyd}
\begin{pmatrix} \dot x \\ \dot y \end{pmatrix}
= \frac{H_i}{2(D-1)} \left[ 2y+(D-2)(D-3)
x^2+2(D-2) x(1+y) \right]
\begin{pmatrix} (D-4) x+ 3 \\ (D-4) (y-1) \end{pmatrix}
\ee This flow has one fixed point at
$O=(x=3/(4-D),y=1)$ and a curve of fixed points
where $2y+(D-2)(D-3) x^2+2(D-2) x(1+y) = 0$. This
line in fact bounds the physical region where
$\rho>0$, the latter containing the fixed point.
Trajectories are thus straight lines in   the
$(x,y)$ plane, which pass through the fixed point
$O$, \be \label{cons} \left( \frac{H_k}{H_i} - 1
\right) = \mu \left( \frac{H_j}{H_i} -
\frac{3}{4-D} \right) \ee where the slope $\mu$
depends on the initial conditions. Substituting
back into (\ref{dife}), one easily computes the
energy density and the time  derivative of the
$i$-th Hubble parameter when  the bounce $H_i=0$
occurs:
 \be
\dot H_i = -\frac{(D-2)(D-4)(2\mu +D-3)}{2(D-1)}
H_j^2\ ,\quad \rho = \frac12 (D-2) (2\mu+D-3) H_j^2
\ee A bounce for direction $i$ in units of the
eleven-dimensional frame therefore takes place for
any initial condition such that $2\mu+D-3>0$  and
 $2<D<4$. A representative trajectory is shown of
Figure (\ref{bfig}), exhibiting a bounce in the $i$
direction. At late times, $(x,y)$ 
approaches the fixed point $O$, so that $H_j \sim 3
H_i/(4-D)$, $H_k \sim H_i$. The radii \be a_i \sim
a_k \sim t^{\frac{2(4-D)}{3(D-2)}}\ ,\quad a_j \sim
t^{\frac{2}{D-2}} \ee thus continue to expand
infinitely.

In order to describe the physics in the string
frame, recall that $1/l_s^2 = a_k/l_M^3$ and
$g_s=(a_k/l_M)^{3/2}$, so that the size of
directions $i$ and $j$ in string units are $a_i
a_k^{1/2}$ and $a_j a_k^{1/2}$, respectively. The
dilaton thus appears to be continuously growing to
infinity at late times, while the direction $i$
undergoes a bounce in the string frame as well,
much before the bounce in the M-theory takes place.

It should however  be noted that the class of
bouncing solutions that we found cannot be smoothly
connected to a Milne geometry in the past, since it
would require an initial condition $(|H_j|,
|H_k|)\ll |H_i|<1$ outside the basin of attraction
for which the bounce occurs.

\subsubsection{Condensation of Long Strings in the
Rindler Regions}

Finally, we would like to discuss the effect of the
condensation of winding strings from the point of
view of an observer in the Rindler region. The
strings are now long strings stretching from radial
infinity towards the origin, and are expected to
deform the geometry. The Einstein equations can be
obtained simply by analytic continuation $T\to ir$,
$a_1(T) \to i b(r)$, leading to a metric and
energy-momentum tensor \be
ds^2=dr^2-b^2(r)d\eta^2+c^2(r) dx_j^2\ ,\quad
T_{rr}=-\frac{{\cal T}_2}{c^{D-1}},\ \
T_{\eta\eta}={\cal T}_2 \frac{b^2}{c^{D-1}} \ee The
main point to observe is a change of sign in the
equation of motion (\ref{difet}) for the volume of
space-time $b c^{D-1}$, \be H' = -H^2  +
\frac{D+1}{D-1} p \ ,\quad p=\frac{{\cal
T}_2}{c^{D-1}} \ee where now $H=(b'/b) + (D-1)
(c'/c)$ and the prime denotes the radial derivative
$d/dr$. The effect of long strings in the Rindler
region is thus to precipitate the appearance of the
singularity $H\to -\infty$. By analytic
continuation, this is consistent with the
observation that short strings in the Milne region
tend to delay the appearance of the singularity.

\subsection{Deformations of Misner Space}
In the previous subsections, we have shown that a
bounce could indeed be induced by the condensation
of winding strings, at least for a low transverse
dimensionality. In this Section, we will postulate
that the Milne geometry is already resolved by the
condensation of winding strings and some other
mechanism, and study the production rate of winding
strings in more general geometries, \be \label{mf}
ds^2 = - dT^2 + a^2(T) d\theta^2 +  dx_i^2 \ee Of
course, unless $a(T)$ is linear in $T$, such a
metric is not a solution of string theory at
tree-level, but we expect that (\ref{mf}) is a
mean-field  description should become on-shell upon
including a time-dependent dilaton profile or
Fischler-Susskind counterterms, to account for
one-loop back-reaction. We assume that $a^2(T)$ is
even under time reversal, and $a(T)$ is analytic on
the real $T$ axis. The metric can then be continued
into a Rindler-like regions with time-independent
metric \be ds^2 = dr^2 - b^2(r) d\eta^2 +  dx_i^2
\ee with $a(ir)=i b(r)$ is a real and  even
function of $r$. Note that if $a(T)$ is bounded,
$b(r)$ will diverge at some point; this is for
example the case of the two-dimensional black hole
geometry \cite{DVV,Craps:2002ii}, \be a(T) = \tanh(
\beta T)\ , \quad b(r) = \tan (\beta r) \ee On the
other hand, even if $a(T)$ is analytic on the real
axis, it may have cuts on the imaginary axis, as
for example in \be \label{resmil} a(T) =
\sqrt{(\beta T)^2+\Delta^2} \ ,\quad b(r) =
\sqrt{(\beta r)^2 -\Delta^2} \ee with $\Delta>0$ real.
In this case, one should retain only the
disconnected regions $r<-\Delta$ and $r>\Delta$, which
have Lorentzian signature. While the Milne region
is now regular, the singularities at $r=\pm\Delta$ in
the Rindler regions are now genuine curvature
singularities (not to mention the deficit angle
introduced by the compactness of
$\theta$)\footnote{In  order for the Rindler
regions to be smoothly capped off  at $r=\Delta$, one
would need $b(r) \propto (r-\Delta) \propto
(r^2+\Delta^2)$, which would imply a Euclidean
signature after continuing to the  Milne region.}.
This is the geometry which seems to be favored by
our qualitative analysis in the previous
subsection.

One may also envisage \be \label{resrin} a(T) =
\sqrt{(\beta T)^2-\Delta^2} \ ,\quad b(r) = \sqrt{r^2
+ \Delta^2} \ee which does not  satisfy the
analyticity condition on the real axis. The two
cosmological regions are now disconnected and
singular, while the two Rindler regions are
connected  by a regular throat.

\subsubsection{Milne Regions}
\label{ssm} Let us first consider the dynamics of
winding strings in the Milne regions. We choose a
gauge such that $\eta(\tau,\sigma)=w \sigma$ and
restrict ourselves to configurations where the time
coordinate $T(\tau)$ is a function of world-sheet
time only. The world-sheet Hamiltonian for $T$ is
then given by a simple generalization of
\eqref{schror}: \be \label{genm} L_0+\bar L_0   =
\frac{1}{a(T)} \pa_T~ a(T)~ \pa_T  + \frac14 w^2
a^2(T) +  \frac{J^2}{a^2(T)} \ee We recognize in
\eqref{genl} the standard $p_L^2+p_R^2$
contribution to $L_0+\bar L_0$ for a compact boson
of radius $a(r)$. For $w=0$, this is simply the
Laplacian in the deformed geometry \eqref{mf}. This
operator has the wrong sign for the kinetic term,
but since we are only interested in states with a
fixed eigenvalue $-\mu^2= -(M^2+\tilde M^2)/2$, we
may simply consider its opposite. In order to
obtain a canonical kinetic term, it is useful to
change variables to $x=\int dT/a(T)$, and rewrite
the opposite of \eqref{genm} as a Schr\"odinger
equation $-\pa_x^2 + V(x) = 0$ with potential \be
V(x) = - \frac14 w^2 a^4(T) - \mu^2 a^2(T) - J^2 =
\frac{M^2 \tM^2}{w ^2} - \left(
\frac{M^2+\tM^2}{2w} + \frac{w}{2} a^2(T) \right)^2
\ee This potential is always negative, and has a
maximum when $a'(T)=0$; it is unbounded from below
if the radius $a(T)$ decompactifies at
$T=\pm\infty$ as in the Misner universe case. The
potential $V(x)$ is finite even at singular points
where $a(T)=0$. Singular points $T_0$ where $a(T)$
vanishes at least linearly are at infinite distance
in the $x$  metric, leading to disconnected
Schr\"odinger problems on the real $x$-axis for
each of the  regions $T<T_0$ and $T>T_0$; the
appropriate continuation between the two requires
evolving the wave function from the ``past Milne
region'' $T<T_0$ into the Rindler patches, and then
back into the ``future Milne region'' $T>T_0$.
Points where $a(T)$ vanishes less than linearly on
the other hand lead to a smooth Schr\"odinger
problem, where trajectories with vanishing mass and
angular momentum barely graze the top of the
potential.

Let us now assume that the production of winding
strings, or  some other mechanism, has resolved the
cosmological singularity and leads to a nowhere
vanishing profile $a(T)$, as in \eqref{resmil}
above. We are thus in a ``scattering above a
potential barrier'' situation (see Figure \ref{misnerdef}, left).
A twisted string
coming in from $T=-\infty$ will  mostly propagate
towards $T=+\infty$; part of its wave function
however will be reflected back towards $T=-\infty$:
as  in the Schwinger effect, this can be
interpreted as stimulated annihilation of the
twisted string. The reflection coefficient can be
computing by approximating the potential around
$x=0$ by an inverted harmonic oscillator, \be V(x)
= V_0 - \frac12 k x^2 + O(x^4)\ ,\quad k= \beta^2
\Delta^2 (2\mu^2  + \Delta^2 w^2) \ee leading to a
reflection coefficient (e.g. \cite{Landau},
chap.50, ex 4) \be {\cal R} \sim \frac{1}{1+e^{-2\pi
V_0/\sqrt{k}}} \ee This is also the spontaneous production rate
of twisted strings. At large winding
number, the production rate is thus exponentially
suppressed as $R\sim \exp(-  \pi \frac{w
\Delta^2}{2\beta} )$, in marked improvement from the
unresolved Misner geometry. Removing the regulator
$\Delta$ at finite $w$ also leads to a vanishingly
small reflection coefficient ${\cal R}\sim \exp(- \pi
J^2 \sqrt{2} /\beta\Delta\mu)$, as expected in the
Misner limit.

Finally, we note that  the classical  potential in
the $T$  variable, $V(x)/a^2(T)$, has only a local
minimum at $T=0$, while the true maxima  lie at
$a^2(T)=2|j|/w$ whenever $w<2|j|/\Delta^2$. The
physical meaning of  these  off-shell states is
unclear.

\FIGURE{ \label{misnerdef}
\hfill \epsfig{file=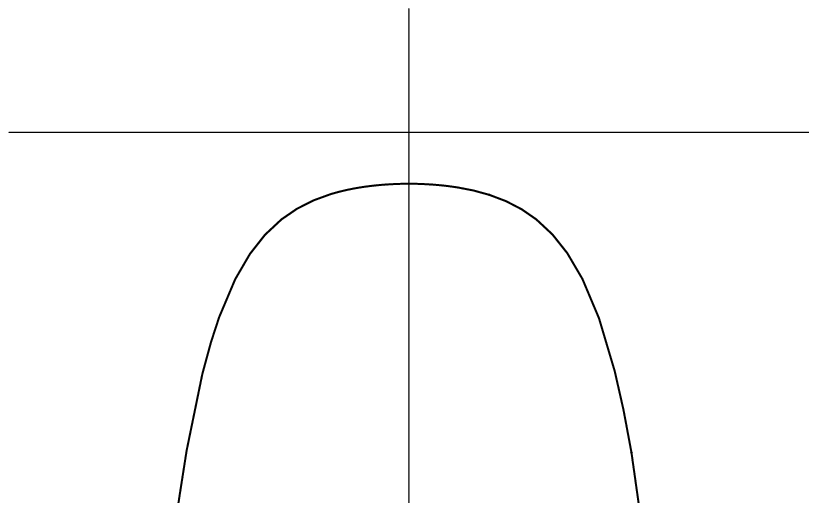,height=4cm} \hfill
\epsfig{file=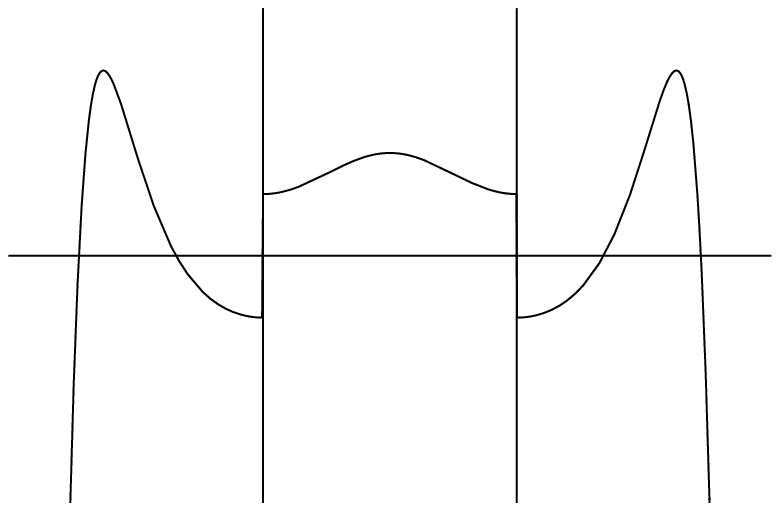,height=4cm} \hfill
\caption{Potential $V$ controling the production of winding strings in the
deformed geometry \eqref{resmil}. Left: potential for a
winding string  in  the  Milne  region, plotted
w.r.t. to the flat coordinate $x$. Right: potential
for a winding string  in  the
Rindler region, plotted w.r.t. to the flat
coordinate $y$. In either case, the horizontal line
denotes the (physical) zero energy level. In the
Rindler case, the intermediate region has
Euclidean signature.} }


\subsubsection{Rindler Regions}
\label{ssr}

Let us now turn to the dynamics of winding strings
in the Rindler region. Restricting to
configurations with $\eta(\tau,\sigma)=w \sigma$
and $r(\tau)$ is a function of world-sheet time
only, the world-sheet Hamiltonian for $r$ is then
given by a simple generalization of \eqref{schror}:
\be \label{genl} L_0+\bar L_0  = - \frac{1}{b(r)}
\pa_r~b(r)~ \pa_r - \frac14 w^2 b^2(r)  -
\frac{J^2}{b^2(r)}\ , \ee equated to $-\mu^2=
-(M^2+\tilde M^2)/2$ by the Virasoro condition.
Again, it is useful to change variables to $y=\int
dr/b(r)$, leading to a Schr\"odinger equation
$-\pa_y^2 +  V(y)=0$ with potential \be
\label{schrorrd} V(y) = - \frac14 w^2 b^4(r) +
\mu^2 b^2(r) - J^2 = \frac{M^2 \tM^2}{w ^2} -
\left( \frac{M^2+\tM^2}{2w} - \frac{w}{2} b^2(r)
\right)^2 \ee Despite the apparent singularity of
the wave equation \eqref{genl} when $b(r)=0$, the
Schr\"odinger equation based on the potential
$V(y)$ is in fact regular (see \cite{Giveon} for an analogous
statement in black hole geometries). If $b(r)$ vanishes
linearly or faster at $r=r_0$, as in the Misner
case, the proper distance between the two sides of
the singularity is infinite in the $y$ coordinate.
In contrast to the Milne regions, these two regions
are causally disconnected and the Hilbert space
factorizes into a tensor  product of states on each
side. As we have seen in Section 3, on either side
of the singularity, the particles can tunnel
between infinity and $r=r_0$, leading to production
of pairs of long and short strings.

Let us now investigate the case where $b(r)$ is
deformed as in \eqref{resmil} with $\Delta^2>0$ (see Figure \ref{misnerdef},
right). The
vanishing locus of $b(r)$ now splits into splits
into two points at $r=\pm \Delta/\beta$, which are at
finite distance in the $y$ coordinate. For
$r>r_0=\Delta/\beta$, one may choose
$r=(\Delta/\beta)\cosh \beta y$ with $y>0$. The
potential has a quadratic minimum with negative
energy at $y=0$, \be V = -J^2 + (\Delta \beta \mu)^2
y^2 + O(y^2)\ ,\quad y>0 \ee then reaches a maximum
with positive energy at $y_1$ given by $\sinh y_1 =
\sqrt{M^2+\tM^2}/(w\Delta)$, such that \be V =
\frac{\mu^4}{w^2} - J^2 -
\frac{2\beta^2\mu^2(2\mu^2+\Delta^2 w^2)}{w^2}
(y-y_1)^2 + O[ (y-y_1)^3] \ee before dropping to
negative values at large $y$. There is thus
tunneling between the $y<y_1$ and $y>y_1$ regions,
just as in the unresolved Misner geometry. The
tunneling rate can be computed as before by
approximating the local maximum with an inverted
harmonic oscillator. In the large $w$ limit, it is
easy to see that the reflection rate reaches a
finite value, \be R \sim \frac{1}{1+\exp\left[ \pi
\frac{J^2 \sqrt{2}}{\beta\Delta \mu} \right]} \ee
which goes to zero when the regulator $\Delta$ is
removed.

However, the point $y=0$ is now at finite distance,
and one needs to enforce boundary conditions there.
A natural prescription is to continue solving the
wave equation, now in the Euclidean region
$-\Delta/\beta< r < \Delta/\beta$. More precisely, in
the region where $b(r)$ is imaginary we define the
real coordinate $dy=i dr/ b(r)$. The wave equation
can now be written as a Schr\"odinger equation
$+\pa_y^2 +  V(y)=0$, with now
$r=(\Delta/\beta)\cos(\beta y)$. In other words, the
potential governing the $y$ evolution in the
$-\Delta/\beta< r < \Delta/\beta$ is now $-V$. Since it
is negative, we are again in a classically
forbidden region. Finally, we revert to Lorentzian
evolution in the $r<-\Delta/\beta$ region,
corresponding to the left Rindler patch.
Altogether, we thus find an exponentially
suppressed coupling between the right and left
Rindler patches, through an intermediate region of
Euclidean signature. The resulting production rates
of long and short strings in each of the Rindler
regions can be computed in a reliable way by
standard WKB techniques.

If on the other hand the deformation corresponds to
\eqref{resrin}, the wave equation becomes regular
for all values of $y$, and allows propagation
between $y>0$ and $y<0$, covering the whole real
$r$ axis. This situation is analogous to the Milne
region discussed in Section \ref{ssm}. The turning
points at zero energy lie at $b^2=(M  \pm
\tM)^2/(w^2)$ and disappear at large $w$,  where
the motion  between $y>0$ and $y<0$ becomes
classically allowed. There is nevertheless pair
production in this case as well, which can be
computed as above by approximating the potential by
an inverted harmonic well.

\section{Conclusion}
In this paper, we have discussed the behavior of
closed strings in Misner space, a toy example of a cosmological
singularity where string theory can in principle be solved.
By studying semiclassically the tree-level spectrum and
one-loop amplitude, we have found several features which we believe
should be of more general validity:
\begin{itemize}

\item[(i)] The usual divergences in the energy-momentum tensor in
field theory are more acute in string theory, due to the existence
of Regge trajectories with arbitrarily high spins. On the other hand,
local expectation values of the energy-momentum tensor are not
directly observable, and the integrated free energy may still be finite.

\item[(ii)]
Winding string production can be understood at a semi-classical level
as tunneling under the barrier in regions with compact time, or scattering
over the barrier in cosmological regions. In general, it can be computed
as a tree-level two-point function in an appropriate basis depending
on the choice of vacuum. In our free field example, it just reduces to the
overlap of the zero-mode wave functions.

\item[(iii)] In the presence of CTC, there may exist closed string
configurations wrapping the
time-like direction. Such strings are better viewed as possibly infinite
static stretched strings. Due to the peculiar causal structure of the
induced metric on the world-sheet, it may be more appropriate to quantize them
with respect to the world-sheet compact coordinate $\sigma$ rather than
$\tau$.

\item[(iv)] Long strings living the whiskers and short strings living in the
cosmological region have a non-trivial two-point function. It is therefore
not legitimate to truncate the space to the cosmological region only.

\item[(v)] In the presence of a cosmological singularity with compact
transverse space, we expect a production of
winding states (strings or D-branes). Their effect on the geometry should
be analogous to that of a positive cosmological constant. If sufficient,
it may prevent the instabilities towards large black hole formation.

\item[(vi)] In addition to the divergences appearing in
the energy-momentum tensor of untwisted fields, the production
rate for winding strings diverges at large winding number.
It should be possible to determine the resolved geometry
self-consistently so that the divergences at one-loop cancel
those at tree-level. Hopefully, there should exist a regime
of parameters where higher loops can consistently be neglected.

\end{itemize}
It is our hope that the Lorentzian
orbifold can teach us some of the new physics which will be
needed to understand gravitational back-reaction in the vicinity
of cosmological singularities.

\vskip 1cm

{\it Acknowledgments}: We are grateful to I.
Affleck, O. Bergman, R. Brustein, B. Craps, B. Durin, A.
Konechny, R. Parentani, D. Reichmann, K. Schleich,
G. Semenoff, K. Skenderis, A. Strominger, M. van
Raamsdonk and E. Verlinde for useful discussions
and comments. The three authors are indebted to the
Aspen Center for Physics for providing a
stimulating atmosphere which lead to this
collaboration. B. P.  and M . R. are grateful to
the Weizmann Institute  for hospitality during part
of this work. M.B. is grateful to KITP and Ecole
Polytechnique for their hospitality during part of
this work. The work of M.B. is supported by Israeli
Academy of Science Centers of Excellence program,
by the Minerva Foundation, by the Harold Blumenstein Foundation and by EEC
grant RTN-2000-001122. The work of M.R. is supported by
NSERC discovery grant.


\begin{thebibliography}{00}


\bibitem{Brandenberger:1988aj}
R.~H.~Brandenberger and C.~Vafa, ``Superstrings In
The Early Universe,'' Nucl.\ Phys.\ B {\bf 316}
(1989) 391.

\bibitem{Lawrence}
A.~E.~Lawrence and E.~J.~Martinec,
``String field theory in curved spacetime and the resolution of spacelike
Class.\ Quant.\ Grav.\  {\bf 13} (1996) 63
[arXiv:hep-th/9509149].

\bibitem{Gubser}
S.~S.~Gubser,
``String production at the level of effective field theory,''
Phys.\ Rev.\ D {\bf 69} (2004) 123507
[arXiv:hep-th/0305099].




\bibitem{Kabat}
A.~J.~Hamilton, D.~Kabat and M.~K.~Parikh,
``A First-quantized formalism for cosmological particle production,''
JHEP {\bf 0407} (2004) 024
[arXiv:hep-th/0311180].


\bibitem{Nappi:1992kv}
C.~R.~Nappi and E.~Witten, ``A Closed, expanding
universe in string theory,'' Phys.\ Lett.\ B {\bf
293}, 309 (1992) [arXiv:hep-th/9206078].


\bibitem{Khoury:2001bz}
J.~Khoury, B.~A.~Ovrut, N.~Seiberg,
P.~J.~Steinhardt and N.~Turok, ``From big crunch to
big bang,'' Phys.\ Rev.\ D {\bf 65} (2002) 086007
[arXiv:hep-th/0108187].


\bibitem{Nekrasov:2002kf}
N.~A.~Nekrasov, ``Milne universe, tachyons, and
quantum group,'' arXiv:hep-th/0203112.

\bibitem{Elitzur:2002rt}
S.~Elitzur, A.~Giveon, D.~Kutasov and
E.~Rabinovici, ``From big bang to big crunch and
beyond,'' JHEP {\bf 0206}, 017 (2002)
[arXiv:hep-th/0204189];

\bibitem{lms}
H.~Liu, G.~Moore and N.~Seiberg, ``Strings in a
time-dependent orbifold,'' JHEP {\bf 0206}, 045
(2002) [arXiv:hep-th/0204168];
``Strings in time-dependent orbifolds,'' JHEP {\bf
0210}, 031 (2002) [arXiv:hep-th/0206182].




\bibitem{Cornalba:2002fi}
L.~Cornalba and M.~S.~Costa, ``A New Cosmological
Scenario in String Theory,'' Phys.\ Rev.\ D {\bf
66}, 066001 (2002) [arXiv:hep-th/0203031].


\bibitem{Craps:2002ii}
B.~Craps, D.~Kutasov and G.~Rajesh, ``String
propagation in the presence of cosmological
singularities,'' JHEP {\bf 0206}, 053 (2002)
[arXiv:hep-th/0205101];


\bibitem{Fabinger:2002kr}
M.~Fabinger and J.~McGreevy, ``On smooth
time-dependent orbifolds and null singularities,''
JHEP {\bf 0306}, 042 (2003) [arXiv:hep-th/0206196].


\bibitem{Berkooz:2002je}
M.~Berkooz, B.~Craps, D.~Kutasov and G.~Rajesh,
``Comments on cosmological singularities in string theory,''
JHEP {\bf 0303} (2003) 031
[arXiv:hep-th/0212215].


\bibitem{Pioline:2003bs}
M.~Berkooz, and B.~Pioline, ``Strings in an
electric field, and the Milne universe,'' JCAP {\bf
0311} (2003) 007 [arXiv:hep-th/0307280].

\bibitem{Craps:2003ai}
B.~Craps and B.~A.~Ovrut,
``Global fluctuation spectra in big crunch / big bang string vacua,''
Phys.\ Rev.\ D {\bf 69} (2004) 066001
[arXiv:hep-th/0308057].


\bibitem{Misner} C. W. Misner, in {\sl Relativity Theory
    and Astrophysics I: Relativity and Cosmology}, edited by J.\
    Ehlers, Lectures in Applied Mathematics, Vol. 8 (American
    Mathematical Society, Providence, 1967), p. 160.


\bibitem{Horowitz:2002mw}
A.~Lawrence, ``On the instability of 3D null
singularities,'' JHEP {\bf 0211}, 019 (2002)
[arXiv:hep-th/0205288];
G.~T.~Horowitz and J.~Polchinski, ``Instability of
spacelike and null orbifold singularities,'' Phys.\
Rev.\ D {\bf 66}, 103512 (2002)
[arXiv:hep-th/0206228].



\bibitem{Schwinger:nm}
J.~S.~Schwinger, ``On Gauge Invariance And Vacuum
Polarization,'' Phys.\ Rev.\  {\bf 82} (1951) 664.



\bibitem{Grant:1992kj}
J.~D.~Grant, ``Cosmic strings and chronology
protection,'' Phys.\ Rev.\ D {\bf 47} (1993) 2388
[arXiv:hep-th/9209102].

\bibitem{Cooper:1992hw}
F.~Cooper, J.~M.~Eisenberg, Y.~Kluger, E.~Mottola
and B.~Svetitsky, ``Particle production in the
central rapidity region,'' Phys.\ Rev.\ D {\bf 48},
190 (1993) [arXiv:hep-ph/9212206].


\bibitem{birrell}
N.D. Birrell and P.C.W. Davies, Quantum Fields in Curved Space,
Cambridge, UK: Univ. Pr



\bibitem{Cornalba:2002nv}
L.~Cornalba, M.~S.~Costa and C.~Kounnas, ``A
resolution of the cosmological singularity with
orientifolds,'' Nucl.\ Phys.\ B {\bf 637}, 378
(2002) [arXiv:hep-th/0204261].


\bibitem{Cornalba:2003ze}
L.~Cornalba and M.~S.~Costa,
``On the classical stability of orientifold cosmologies,''
Class.\ Quant.\ Grav.\  {\bf 20} (2003) 3969
[arXiv:hep-th/0302137].



\bibitem{Drukker:2003sc}
N.~Drukker, B.~Fiol and J.~Simon,
``Goedel's universe in a supertube shroud,''
Phys.\ Rev.\ Lett.\  {\bf 91} (2003) 231601
[arXiv:hep-th/0306057].



\bibitem{Israel:2003cx}
D.~Israel,
``Quantization of heterotic strings in a Goedel/anti de Sitter spacetime and
chronology protection,''
JHEP {\bf 0401} (2004) 042
[arXiv:hep-th/0310158].



\bibitem{Gabriel:1999yz}
C.~Gabriel and P.~Spindel, ``Quantum charged fields
in Rindler space,'' Annals Phys.\  {\bf 284} (2000)
263 [arXiv:gr-qc/9912016].



\bibitem{bdpr}
M.~Berkooz, B.~Durin, B.~Pioline and D.~Reichmann,
"Closed strings in misner space: stringy fuzziness with a twist",
arXiv: hep-th/0407216.



\bibitem{Strominger:2003fn}
A.~Strominger and T.~Takayanagi,
``Correlators in timelike bulk Liouville theory,''
Adv.\ Theor.\ Math.\ Phys.\  {\bf 7} (2003) 369
[arXiv:hep-th/0303221].


\bibitem{Hawking:1991nk}
S.~W.~Hawking, ``The Chronology protection
conjecture,'' Phys.\ Rev.\ D {\bf 46}, 603 (1992).

\bibitem{elecfield}
E.~S.~Fradkin and A.~A.~Tseytlin, ``Nonlinear
Electrodynamics From Quantized Strings,'' Phys.\
Lett.\ B {\bf 163}, 123 (1985);
A.~Abouelsaood, C.~G.~Callan, C.~R.~Nappi and
S.~A.~Yost, ``Open Strings In Background Gauge
Fields,'' Nucl.\ Phys.\ B {\bf 280}, 599 (1987).
C.~P.~Burgess, ``Open String Instability In
Background Electric Fields,'' Nucl.\ Phys.\ B {\bf
294}, 427 (1987).;



\bibitem{Bachas:bh}
C.~Bachas and M.~Porrati, ``Pair Creation Of Open
Strings In An Electric Field,'' Phys.\ Lett.\ B
{\bf 296}, 77 (1992) [arXiv:hep-th/9209032].

\bibitem{affleck}
I.~K.~Affleck and N.~S.~Manton, ``Monopole Pair
Production In A Magnetic Field,'' Nucl.\ Phys.\ B
{\bf 194}, 38 (1982); I.~K.~Affleck, O.~Alvarez and
N.~S.~Manton, ``Pair Production At Strong Coupling
In Weak External Fields,'' Nucl.\ Phys.\ B {\bf
197}, 509 (1982).


\bibitem{nielsen}
N.~K.~Nielsen and P.~Olesen,
``An Unstable Yang-Mills Field Mode,''
Nucl.\ Phys.\ B {\bf 144} (1978) 376.

\bibitem{David:2002km}
J.~R.~David,
``Unstable magnetic fluxes in heterotic string theory,''
JHEP {\bf 0209} (2002) 006
[arXiv:hep-th/0208011].

\bibitem{Hiscock:vq}
W.~A.~Hiscock and D.~A.~Konkowski, ``Quantum Vacuum
Energy In Taub - Nut (Newman-Unti-Tamburino) Type
Cosmologies,'' Phys.\ Rev.\ D {\bf 26} (1982) 1225.





\bibitem{sushkov}
 S.~V.~Sushkov,
 ``Chronology protection and quantized fields: Complex automorphic scalar
Class.\ Quant.\ Grav.\  {\bf 14} (1997) 523
[arXiv:gr-qc/9509056].



\bibitem{Maldacena:2000kv}
J.~M.~Maldacena, H.~Ooguri and J.~Son, ``Strings in
AdS(3) and the SL(2,R) WZW model. II: Euclidean
black hole,'' J.\ Math.\ Phys.\ {\bf 42}, 2961
(2001) [arXiv:hep-th/0005183].



 \bibitem{BKL}
E.~M.~Lifshitz and I.~M.~Khalatnikov,
``Investigations In Relativistic Cosmology,'' Adv.\
Phys.\  {\bf 12}, 185 (1963);
V.~A.~Belinsky, I.~M.~Khalatnikov and
E.~M.~Lifshitz, ``Oscillatory Approach To A
Singular Point In The Relativistic Cosmology,''
Adv.\ Phys.\  {\bf 19}, 525 (1970);
V.~a.~Belinsky, I.~m.~Khalatnikov and
E.~m.~Lifshitz, ``A General Solution Of The
Einstein Equations With A Time Singularity,'' Adv.\
Phys.\  {\bf 31}, 639 (1982).


\bibitem{Tseytlin:1991xk}
A.~A.~Tseytlin and C.~Vafa, ``Elements of string
cosmology,'' Nucl.\ Phys.\ B {\bf 372} (1992) 443
[arXiv:hep-th/9109048].


\bibitem{Molina-Paris:1998tx}
C.~Molina-Paris and M.~Visser,
``Minimal conditions for the creation of a Friedman-Robertson-Walker  universe
Phys.\ Lett.\ B {\bf 455}, 90 (1999)
[arXiv:gr-qc/9810023].


\bibitem{DVV}
R.~Dijkgraaf, H.~Verlinde and E.~Verlinde,
``String propagation in a black hole geometry,''
Nucl.\ Phys.\ B {\bf 371} (1992) 269.




\bibitem{Landau}
L.~Landau and E.~Lifschitz, {\it Mechanique Quantique}, Editions Mir,
Moscou; Section 50, Exercice 4.


\bibitem{Giveon}
A.~Giveon, B.~Kol, A.~Ori and A.~Sever,
 ``On the resolution of the time-like singularities in Reissner-Nordstroem and
negative-mass Schwarzschild,''
arXiv:hep-th/0401209.


\end{thebibliography}
\end{document}